\documentclass[nofootinbib,superscriptaddress,aps,preprint,prc]{revtex4}
\usepackage[utf8]{inputenc}
\usepackage{amsmath}
\usepackage{mathrsfs}
\usepackage{amssymb}
\usepackage{bbm}
\usepackage{graphicx}
\usepackage{graphics}
\usepackage[small,compact]{titlesec}
\usepackage{xspace}
\usepackage{colordvi}
\usepackage{color}
\usepackage{units}
\usepackage{stackrel}
\usepackage{hyperref}
\usepackage{array}
\usepackage{slashed}
\usepackage{ulem}

\renewcommand{\emph}[1]{\textit{#1}}

\newcommand{\vect}[1]{\vec{\mathbf{#1}}}

\newcommand{\EFT}{$\mathrm{EFT}(\slashed{\pi})$\xspace}

\newcommand{\comment}[1]{}

\newcommand{\G}{\mathcal{G}}
\newcommand{\Gb}{\boldsymbol{\mathcal{G}}}

\makeatletter

\newcommand{\Rmnum}[1]{\expandafter\@slowromancap\romannumeral #1@}
\newcommand{\vast}{\bBigg@{4}}
\newcommand{\Vast}{\bBigg@{5}}

\newcolumntype{P}[1]{>{\centering\arraybackslash}p{#1}}

\makeatother

\begin{document}

\title{Charge and Matter Form Factors of Two-Neutron Halo Nuclei in Halo Effective Field Theory at Next-to-leading-order}

\author{Jared Vanasse}
\email{vanasse@.ohio.edu}

\affiliation{Department of Physics and Astronomy Ohio University, Athens OH 45701, USA
}

\date{\today}

\begin{abstract}
Using halo effective field theory (EFT), an expansion in $R_{core}/R_{halo}$, where $R_{core}$ is the radius of the core and $R_{halo}$ the radius of the halo nucleus, we calculate the charge and neutron form factors of the two-neutron halo nuclei $^{11}$Li, $^{14}$Be, and $^{22}$C to next-to-leading-order (NLO) by treating them as an effective three-body system.  From the form factors we extract the point charge and point matter radii, inter-neutron distance, and neutron opening angle.  Agreement is found with existing experimental extractions.  Results are given for the point charge and point matter radii for arbitrary neutron core scattering effective range, $\rho_{cn}$, that can be used for predictions once $\rho_{cn}$ is measured.  Estimates for $\rho_{cn}$ are also used to make NLO predictions.  Finally, our point charge radii are compared to other halo-EFT predictions, and setting the core mass equal to the neutron mass our point charge radius is found to agree with an analytical prediction in the unitary limit.

\end{abstract}


\maketitle

\newpage

\section{Introduction}

When probing distance scales much larger than the scale of the underlying interaction, $r$, interactions can be approximated in a series of contact interactions known as short-range effective field theory (EFT).  The wide applicability of this formalism to low energy systems such as cold atoms, low energy few-nucleon systems, and halo nuclei is known as universality~\cite{Braaten:2004rn}.  Short-range EFT is an expansion in  $M_{low}/M_{high}$, where $M_{high}\sim 1/r$ sets the scale of physics not explicitly included, and $M_{low}\sim Q$, with $Q$ a typical momentum scale in the problem.  For most systems of interest in short-range EFT it is found that the two-body $S$-wave scattering length, a, scales unnaturally ($M_{low}\sim 1/a <M_{high}$).  This requires leading-order (LO) interactions to be treated non-perturbatively leading to the creation of shallow two-body bound states~\cite{Kaplan:1998tg,Kaplan:1998we}.  Higher-order range interactions are added perturbatively on top of the LO results in an expansion in powers of $r/a$.  In this work we focus on two-neutron halo nuclei through the short-range EFT known as halo-EFT, however, via universality the methods and results are equally applicable to cold atom systems and low energy few-nucleon systems using pionless EFT (\EFT).

Halo nuclei found along the nuclear drip lines are characterized by a core of size, $R_{core}\sim 1/M_{high}$, and loosely bound valence nucleons giving the size of the halo nucleus, $R_{halo}\sim 1/M_{low}$, such that $R_{core}\ll R_{halo}$.  Halo-EFT takes advantage of these disparate scales by expanding in powers of $(R_{core}/R_{halo})$.  In halo-EFT the core is treated as a fundamental degree of freedom with no internal structure.  Breakdown of this description occurs at energy scales $M_{high}\sim E_{C}^{*}$ or $M_{high}\sim B_{c-n}$, where $E_{C}^{*}$ is the first excited state energy of the core and $B_{c-n}$ the one neutron separation energy of the core.  At these energies the core can no longer be treated as a fundamental degree of freedom.  The typical momentum scale of the halo nucleus is given by its binding energy $B_{halo}\sim M_{low}$.  In addition to offering a systematically improvable method for calculating properties of halo nuclei, halo-EFT also allows for estimation of theoretical errors.

In the two-body sector halo-EFT was introduced to study $p$-wave resonance interactions in $n\alpha$ scattering~\cite{Bertulani:2002sz,Bedaque:2003wa}.  It has also been used to investigate  properties of the one neutron halo nuclei such as $^{8}$Li~\cite{Rupak:2011nk,Fernando:2011ts}, $^{15}$C~\cite{Rupak:2012cr} and $^{11}$Be and $^{19}$C~\cite{Fernando:2015jyd}.  Investigation into the possibility of excited Efimov states of two-neutron halo nuclei with dominant $S$-wave interactions was carried out in Ref.~\cite{Canham:2008jd} at LO.  This work also considered point charge and point matter radii of two-neutron halo nuclei, and was later extended to next-to-leading-order (NLO)~\cite{Canham:2009xg} by including range corrections.  However, the NLO calculation was not strictly perturbative as it resummed range corrections to all orders.  This calculation had all the necessary contributions to NLO, but contained an infinite subset of higher order terms.  The two-neutron halo $^{6}$He was considered in Refs.~\cite{Rotureau:2012yu,Ji:2014wta} by including two-body resonant $P$-wave interactions.  Examination of the two-neutron halo $^{22}$C matter radius was carried out to LO in Ref.~\cite{Acharya:2013aea}, and the charge radii of the two-neutron halos $^{11}$Li, $^{14}$Be, and $^{22}$C were calculated by Hagen et \emph{al.}~\cite{Hagen:2013xga} at LO.

Building upon the work of Hagen et \emph{al.}, Vanasse~\cite{Vanasse:2015fph} calculated the triton charge radius in \EFT to next-to-next-to-leading-order (NNLO).  In this work we will calculate the charge and matter form factors and radii of the two-neutron halos $^{11}$Li, $^{14}$Be, and $^{22}$C to NLO by adding range corrections perturbatively.  Note, this differs from the work of Canham and Hammer~\cite{Canham:2009xg} in which range corrections are summed to all orders.  In addition to showing NLO results we also demonstrate that the point charge radii results of Hagen et \emph{al.}~\cite{Hagen:2013xga} are incorrect, most likely due to a wrong factor in front of a single term.  Our analytical functions at LO for the charge form factor nearly agree with those of Hagen et \emph{al.}~except in one instance.  Using the slightly modified functions of Hagen et \emph{al.}~we are able to reproduce their results, however, we find in the unitary and equal mass limit that they do not agree with an analytical solution for the point charge radius~\cite{Braaten:2004rn}.  Using our form for the analytical functions we obtain the correct point charge radius in the unitary and equal mass limit and also find different point charge radii from Hagen et \emph{al.} for $^{11}$Li, $^{12}$Be, and $^{22}$C.

This work introduces the Lagrangian for halo-EFT is Sec.~\ref{sec:Lag} and interactions in the two-body sector in Sec.~\ref{sec:twobody}.  The trimer vertex function is discussed in Sec.~\ref{sec:threebody} and the formalism for the charge and neutron form factors in Sec.~\ref{sec:formfactor}.  In Sec.~\ref{sec:observable} the basic observables of interest for two-neutron halo nuclei are reviewed.  Sec.~\ref{sec:results} gives the LO and NLO point charge and point matter radii for $^{11}$Li, $^{14}$Be, and $^{22}$C and compares them with available experimental data.  NLO results use naturalness assumptions to estimate the core neutron effective range, $\rho_{cn}$.  The inter-neutron separation and neutron opening angle are also calculated and compared with experimental data. Also given are the NLO corrections to the charge and matter radii for arbitrary $\rho_{cn}$ that can be used to calculate NLO corrections once experimental data is available for $\rho_{cn}$.  Finally, we conclude in Sec.~\ref{sec:conclsuion}.

\section{\label{sec:Lag} Lagrangian and Formalism}

At LO in halo-EFT two-neutron halo nuclei are described by zero range interactions between the $cn$ and $nn$ two-body sub-systems.  NLO adds range correction interactions between the $cn$ and $nn$ sub-systems.  These two-body interactions are encoded in the Lagrangian
\begin{align}
\label{eq:L2}
&\mathcal{L}_{2}=\hat{c}^{\dagger}\left(iD_{0}+\frac{\vect{D}^{2}}{2m_{c}}\right)\hat{c}+\hat{n}^{\dagger}\left(i\partial_{0}+\frac{\vect{\nabla}^{2}}{2m_{n}}\right)\hat{n}+\hat{d}_{0}^{\dagger}\left[\Delta_{0}+w_{0}\left(i\partial_{0}+\frac{\vect{\nabla}^{2}}{4m_{n}}+\frac{\gamma_{nn}^{2}}{m_{n}}\right)\right]\hat{d}_{0}\\\nonumber
&+\hat{d}_{1}^{\dagger}\left[\Delta_{1}+w_{1}\left(iD_{0}+\frac{\vect{D}^{2}}{2(m_{n}+m_{c})}+\frac{\gamma_{cn}^{2}}{2\mu}\right)\right]\hat{d}_{1}-\frac{g_{0}}{\sqrt{8}}\hat{d}_{0}^{\dagger}\hat{n}^{T}i\sigma_{2}\hat{n}-g_{1}\hat{d}_{1}^{\dagger}\hat{n}\hat{c}+\mathrm{H.c.},
\end{align}
where $\hat{c}$ is the core field, $\hat{n}$ the neutron field, and $\hat{d}_{0}$ ($\hat{d}_{1}$) an auxiliary dimer field of the $nn$ ($cn$) system.  The $nn$-dimer, $\hat{d}_{0}$, is not a physical degree of freedom assuming the nonexistence of the di-neutron.\footnote{For work discussing the existence of a di-neutron see Refs.~\cite{Hammer:2014rba,Howell:2016egy}}  Likewise, the $cn$-dimer, $\hat{d}_{1}$,  only corresponds to a physical degree of freedom if the $cn$ system is bound.  Despite dimer fields being unphysical they are still useful in the calculation of bound $cnn$ systems.  The coefficient $g_{0}$ ($g_{1}$) sets the strength of the interaction between the $nn$-dimer and two neutrons ($cn$-dimer and core and neutron).  Gauging the derivatives of the charged core and $cn$-dimer gives the covariant derivative
\begin{equation}
D_{\mu}=\partial_{\mu}+ieZ\hat{A}_{\mu},
\end{equation}
where $Z$ is the number of protons in the core.  To calculate the neutron form factor of $cnn$ systems derivatives acting on neutrons and the $nn$-dimer can be gauged with a ficticious neutron charge not shown in this Lagrangian.  Neutron and charge form factors are both necessary to extract the matter radii of halo nuclei.  The mass of the core and neutron are given by $m_{c}$ and $m_{n}$ respectively, while their reduced mass is given by 
\begin{equation}
\mu=\frac{m_{n}m_{c}}{m_{n}+m_{c}}.
\end{equation}
$\Delta_{0}$ ($\Delta_{1}$) is the bare $nn$-dimer ($cn$-dimer) propagator, and $\gamma_{0}$ ($\gamma_{1}$) the binding momentum of the $nn$ virtual bound state ($cn$ real or virtual bound state).  The parameters $w_{0}$ and $w_{1}$ are proportional to range corrections.  Finally, $i\sigma_{2}$ is a Pauli matrix that projects out the spin-singlet combination of neutrons.  All values of the two-body parameters are given in the next section.

In addition to two-body interactions at LO a three-body interaction must also be included to properly renormalize the three-body system~\cite{Bedaque:1998kg,Bedaque:1998km}.  This is most easily achieved by the introduction of a trimer field $\hat{t}$ that interacts with a core and $nn$-dimer via the Lagrangian~\cite{Bedaque:2002yg,Hagen:2013xga}
\begin{equation}
\label{eq:L3}
\mathcal{L}_{3}=\Omega\hat{t}^{\dagger}\hat{t}-h_{0}\hat{t}^{\dagger}\hat{c}\hat{d}_{0}-h_{1}\hat{t}^{\dagger}\hat{c}\hat{d}_{0}+\mathrm{H.c.}
\end{equation}
The parameter $\Omega$ is the bare trimer propagator, $h_{0}$ is the LO interaction between the trimer, core, and $nn$-dimer, and $h_{1}$ is the NLO correction to $h_{0}$ introduced to avoid refitting at NLO.  Both $h_{0}$ and $h_{1}$ are fit to the $cnn$ bound state energy.  Note, the form for the trimer Lagrangian is not unique~\cite{Hagen:2013xga}.

\section{\label{sec:twobody}Two-Body Systems}

\begin{figure}[hbt]
\includegraphics[width=130mm]{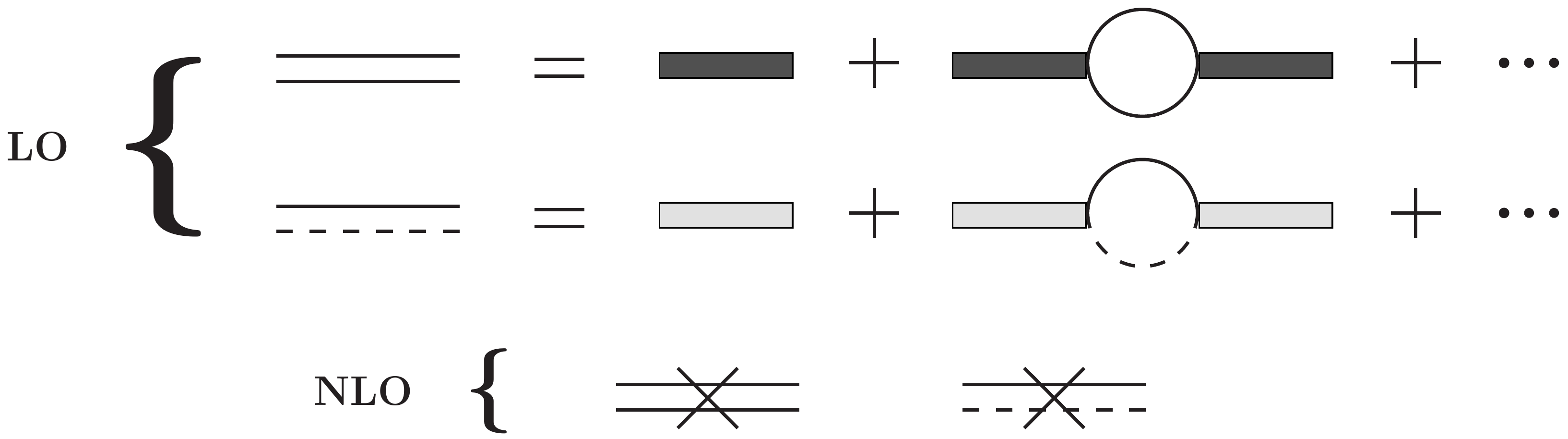}
\caption{\label{fig:HaloDimers}Infinite sum of bubble diagrams that give LO $cn$ and $nn$-dimer propagators. Solid lines are neutrons, dashed lines are cores, light rectangles are bare $cn$-dimer propagators, $i/\Delta_{1}$, and dark rectangles are bare $nn$-dimer propagators, $i/\Delta_{0}$.  The NLO dimer propagators receive range corrections represented by a cross.}
\end{figure}

The LO dimer propagators are given by the infinite sum of diagrams in Fig.~\ref{fig:HaloDimers}.  Solid lines are neutrons, dashed lines the core, the dark rectangle is the bare $nn$-dimer propagator, $i/\Delta_{0}$, and the light rectangle is the bare $cn$-dimer propagator, $i/\Delta_{1}$.  At NLO the dimer propagators receive range corrections represented by crosses in Fig.~\ref{fig:HaloDimers}. 
The infinite sum of diagrams is readily solved via a geometric series yielding the NLO $nn$-dimer propagator
\begin{equation}
\label{eq:nndimer}
D_{nn}(q_{0},q)=\frac{1}{\sqrt{\frac{1}{4}q^{2}-m_{n}q_{0}-i\epsilon}-\gamma_{nn}}\left\{\underbrace{1\vphantom{\frac{Z_{nn}-1}{2\gamma_{nn}}\left(\gamma_{nn}+\sqrt{\frac{2+A}{4A}q^{2}-m_{n}E}\right)}}_{\mathrm{LO}}+\underbrace{\frac{Z_{nn}-1}{2\gamma_{nn}}\left(\gamma_{nn}+\sqrt{\frac{1}{4}q^{2}-m_{n}q_{0}-i\epsilon}\right)}_{\mathrm{NLO}}\right\},
\end{equation}
and NLO $cn$-dimer propagator

\begin{align}
\label{eq:cndimer}
&D_{cn}(q_{0},q)=\frac{1}{\sqrt{\frac{A}{(1+A)^{2}}q^{2}-\frac{2A}{1+A}m_{n}q_{0}-i\epsilon}-\gamma_{cn}}\\\nonumber
&\hspace{4cm}\times\left\{\underbrace{1\vphantom{\frac{Z_{cn}-1}{2\gamma_{cn}}\left(\gamma_{cn}+\sqrt{\frac{A}{(1+A)^{2}}q^{2}-\frac{2A}{1+A}m_{n}q_{0}}\right)}}_{\mathrm{LO}}+\underbrace{\frac{Z_{cn}-1}{2\gamma_{cn}}\left(\gamma_{cn}+\sqrt{\frac{A}{(1+A)^{2}}q^{2}-\frac{2A}{1+A}m_{n}q_{0}-i\epsilon}\right)}_{\mathrm{NLO}}\right\},
\end{align}
%
%\begin{equation}
%D_{cn}(q_{0},q)=\frac{1}{\sqrt{\frac{A}{(1+A)^{2}}q^{2}-\frac{2A}{1+A}m_{n}q_{0}}-\gamma_{cn}}\left\{1+\frac{r_{cn}}{2}\frac{\frac{A}{(1+A)^{2}}q^{2}-\frac{2A}{1+A}m_{n}q_{0}}{\sqrt{\frac{A}{(1+A)^{2}}q^{2}-\frac{2A}{1+A}m_{n}q_{0}}-\gamma_{cn}}\right\}.
%\end{equation}
%
where $A=\frac{m_{c}}{m_{n}}$.  Parameters of the dimer propagators are fit using the $Z$-parametrization~\cite{Phillips:1999hh,Griesshammer:2004pe}, which fits to the pole in the two-body scattering amplitude at LO and to its residue at NLO.  The parameter $\gamma_{nn}$ is fit to the $nn$ virtual bound state momentum, which can be related to the $nn$-scattering length, $a_{nn}$, and effective range, $\rho_{nn}$, via~\cite{Griesshammer:2004pe}
\begin{equation}
\gamma_{nn}=\frac{1}{a_{nn}}+\frac{\rho_{nn}}{2}\gamma_{nn}^{2}+\cdots
\end{equation}
The residue to NLO about the $nn$ virtual bound state pole is given by
\begin{equation}
Z_{nn}=\frac{1}{1-\gamma_{nn}\rho_{nn}}.
\end{equation}
Using the values $a_{nn}=-18.7(6)$~fm~\cite{GonzalezTrotter:1999zz} and $\rho_{nn}=2.75(11)$~fm~\cite{Miller:1990iz} for the $nn$ scattering length and effective range yields the value $\gamma_{nn}=-9.87$~MeV ($Z_{nn}=1.16$) for the $nn$ virtual bound state momentum (residue).

For the $cn$-dimer, $\gamma_{cn}=\mathrm{sign}(B_{1})\sqrt{2\mu |B_{1}|}$, is fit to the $cn$ system ``binding energy", $B_{1}$.  Negative $B_{1}$ values give virtual bound states, and the imaginary part of the binding momentum for such resonant $cn$-states is ignored.  The value of the residue, $Z_{cn}$, about the $cn$ pole is given by
\begin{equation}
Z_{cn}=\frac{1}{1-\gamma_{cn}\rho_{cn}}
\end{equation}
where $\rho_{cn}$ is the effective range for $cn$ scattering.  Unfortunately, experimental determinations of $\rho_{cn}$ are currently unavailable. Therefore, NLO corrections from $Z_{nn}$ and $Z_{cn}$ will be disentangled, and results will be given for arbitrary values of $Z_{cn}$, which can be used to easily determine charge and matter radii once $\rho_{cn}$ is measured.  In addition $\rho_{cn}\sim 1/m_{\pi}=1.4$~fm, will be given a value based on naturalness to make NLO predictions, where $m_{\pi}$ is the pion mass.

Finally the parameters in the two-body Lagrangian are given by~\cite{Griesshammer:2004pe}

\begin{align}
& \Delta_{0}=\gamma_{nn}-\mu_{\mathrm{PDS}},&w_{0}=(Z_{nn}-1)\frac{m_{n}}{2\gamma_{nn}},&&g_{0}^{2}=\frac{4\pi}{m_{n}}\\\nonumber
& \Delta_{1}=\gamma_{cn}-\mu_{\mathrm{PDS}},&w_{1}=(Z_{cn}-1)\frac{\mu}{\gamma_{cn}},&&g_{1}^{2}=\frac{2\pi}{\mu}.
\end{align}
The scale $\mu_{\mathrm{PDS}}$ comes from using dimensional regularization with the power divergence subtraction technique~\cite{Kaplan:1998tg,Kaplan:1998we} for all loop integrals.

%\gamma_{0}=-9.87301735564498
%Z_{0}=1.15954520882204

\section{\label{sec:threebody} Three-Body System}

\begin{figure}[hbt]
\includegraphics[width=110mm]{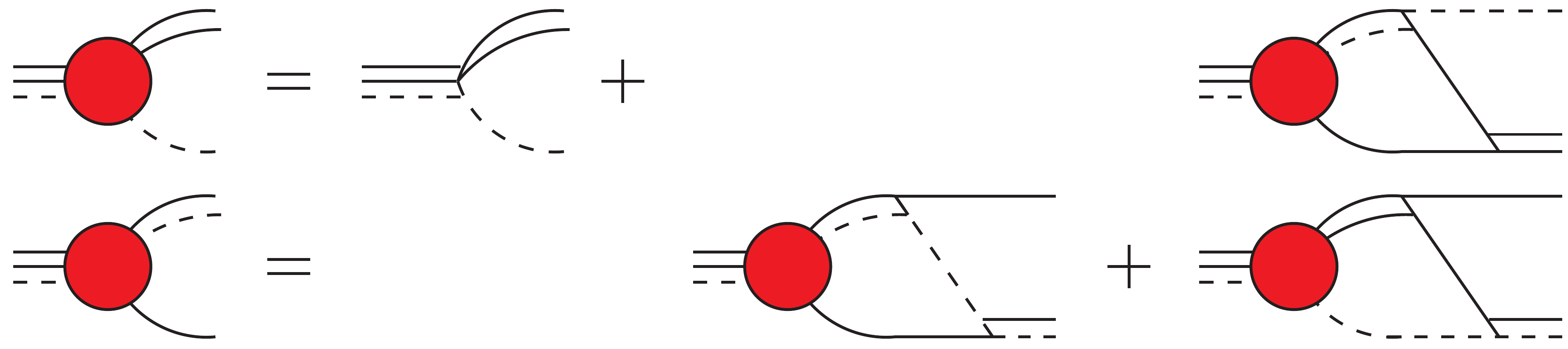}
\caption{\label{fig:HaloLO}Coupled integral equations for LO trimer vertex function.  The trimer field is given by the triple line and the trimer vertex function the the red circle.}
\end{figure}

Calculation of bound state properties of two-neutron halo nuclei requires the three-body wavefunction, which is directly related to the trimer vertex function.  The LO trimer vertex function is given by the coupled integral equations in Fig.~\ref{fig:HaloLO}
, which give the matrix equation
\begin{equation}
\Gb_{0}(E,p)=\mathbf{B}+\Big{[}\mathbf{R}_{0}(E,p,q)\mathbf{D}(E,q)\Big{]}\otimes\Gb_{0}(E,q),
\end{equation}
where the ``$\otimes$" operator is defined by
\begin{equation}
A(q)\otimes B(q)=\frac{1}{2\pi^{2}}\int_{0}^{\Lambda}dq q^{2}A(q)B(q).
\end{equation}
$\Lambda$ is a cutoff used to regulate potential divergences.  Once properly renormalized all physical quantities should have a well defined limit in the limit $\Lambda\to\infty$.  $\mathbf{B}$ the inhomogenous term and $\Gb_{0}(E,p)$ the LO trimer vertex function are both vectors defined by
\begin{equation}
\mathbf{B}=\left(\begin{array}{c}
1 \\[-2mm]
0
\end{array}\right)\quad,\quad
\Gb_{m}(E,p)=\left(\begin{array}{c}
\G_{m,t\to d_{0}c}(E,p)\\[-2mm]
\G_{m,t\to d_{1}n}(E,p)
\end{array}\right),
\end{equation}
where $\G_{m,t\to d_{0}c}(E,p)$ ($\G_{m,t\to d_{1}n}(E,p)$) is the vertex function for a trimer going to a spectator core and $nn$-dimer (spectator neutron and $cn$-dimer).\footnote{Note, that the ``physical" inhomogeneous term should go like $h$ from Eq.~(\ref{eq:L3}).  However, since the normalization of the trimer vertex function is arbitrary the value of one is given to the inhomogeneous term.  Once the trimer vertex function is properly renormalized the scaling will be fixed.}   The subscript ``$m$" refers to the order of the trimer vertex function (i.e. $m=0$ is LO, $m=1$ is NLO, etc...).  The kernel term $\mathbf{R}_{0}(E,p,q)$ is a matrix defined by
\begin{equation}
\mathbf{R}_{0}(E,p,q)=\left(\begin{array}{cc}
R_{00}(E,p,q) & R_{01}(E,p,q)\\
R_{10}(E,p,q) & R_{11}(E,p,q)\\
\end{array}\right),
\end{equation}
where
\begin{equation}
R_{00}(E,p,q)=0,
\end{equation}
\begin{equation}
R_{01}(E,p,q)=\frac{2\sqrt{2}\pi(1\!+\!A)}{A}\frac{1}{pq}Q_{0}\left(\frac{\frac{1+A}{2A}p^{2}+q^{2}-m_{n}E}{pq}\right),
\end{equation}
\begin{equation}
R_{10}(E,p,q)=R_{01}(E,q,p),
\end{equation}
and
\begin{equation}
R_{11}(E,p,q)=\frac{(1\!+\!A)^2\pi}{A}\frac{1}{pq}Q_{0}\left(\frac{\frac{(1+A)}{2}(p^{2}+q^{2})-Am_{n}E}{pq}\right).
\end{equation}
$Q_{0}(a)$ is a Legendre function of the second kind defined by
\begin{equation}
Q_{0}(a)=\frac{1}{2}\ln\left(\frac{1+a}{a-1}\right).
\end{equation}
Finally $\mathbf{D}(E,q)$ is a matrix of LO dimer propagators given by
\begin{equation}
\mathbf{D}(E,q)=\left(\begin{array}{cc}
\bar{D}_{nn}^{(0)}(E,q) & 0\\
0 & \bar{D}_{cn}^{(0)}(E,q) \\
\end{array}\right),
\end{equation}
with
\begin{equation}
\bar{D}_{nn}^{(m)}(E,q)=D_{nn}^{(m)}\left(E-\frac{q^{2}}{2Am_{n}},q\right)\quad,\quad \bar{D}_{cn}^{(m)}(E,q)=\frac{2A}{1+A}D_{cn}^{(m)}\left(E-\frac{q^{2}}{2m_{n}},q\right),
\end{equation}
where the superscript ``($m$)" refers to only the LO (NLO) part of Eqs.~(\ref{eq:nndimer}) and (\ref{eq:cndimer}) for $m=0$ ($m=1$).

\begin{figure}[hbt]
\includegraphics[width=110mm]{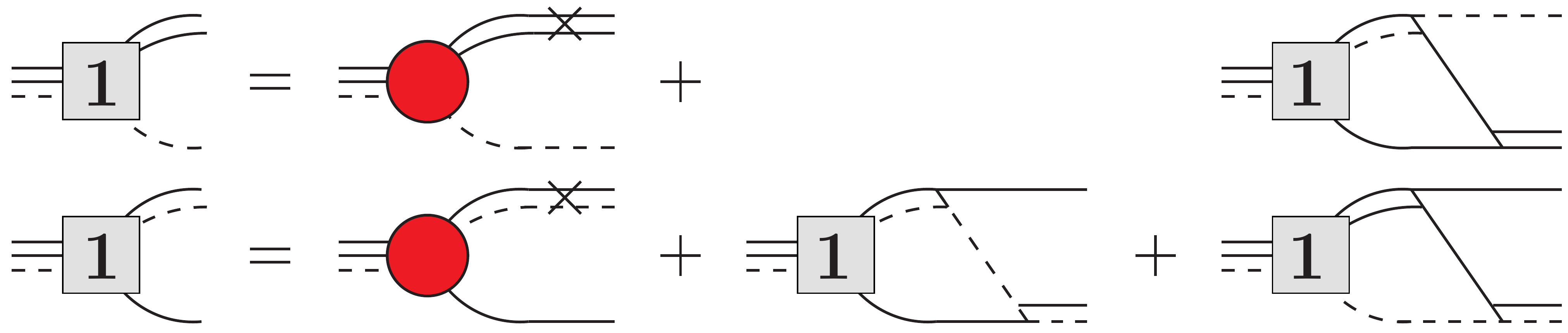}
\caption{\label{fig:HaloNLO}Coupled integral equations for the NLO correction to the trimer vertex function.  The box with a ``1" inside represents the NLO correction to the trimer vertex function.}
\end{figure}

The NLO correction to the trimer vertex function receives range corrections as shown in the coupled integral equations of Fig.~\ref{fig:HaloNLO}
This set of coupled integral equations gives the matrix equation
\begin{equation}
\Gb_{1}(E,p)=\mathbf{R}_{1}(E,p)\Gb_{0}(E,p)+\Big{[}\mathbf{R}_{0}(E,p,q)\mathbf{D}(E,q)\Big{]}\otimes\Gb_{1}(E,q),
\end{equation}
where the matrix $\mathbf{R}_{1}(E,p)$ is 
\begin{equation}
\mathbf{R}_{1}(E,p)=\left(\begin{array}{cc}
\frac{Z_{nn}-1}{2\gamma_{nn}}\left(\gamma_{nn}+\sqrt{\frac{2+A}{4A}p^{2}-m_{n}E} \right) & 0 \\[2 mm]
0 &\frac{Z_{cn}-1}{2\gamma_{cn}}\left(\gamma_{cn}+\sqrt{\frac{A(2+A)}{(1+A)^{2}}p^{2}-\frac{2A}{1+A}m_{n}E}\right)
\end{array}\right).
\end{equation}

\begin{figure}[hbt]
\includegraphics[width=70mm]{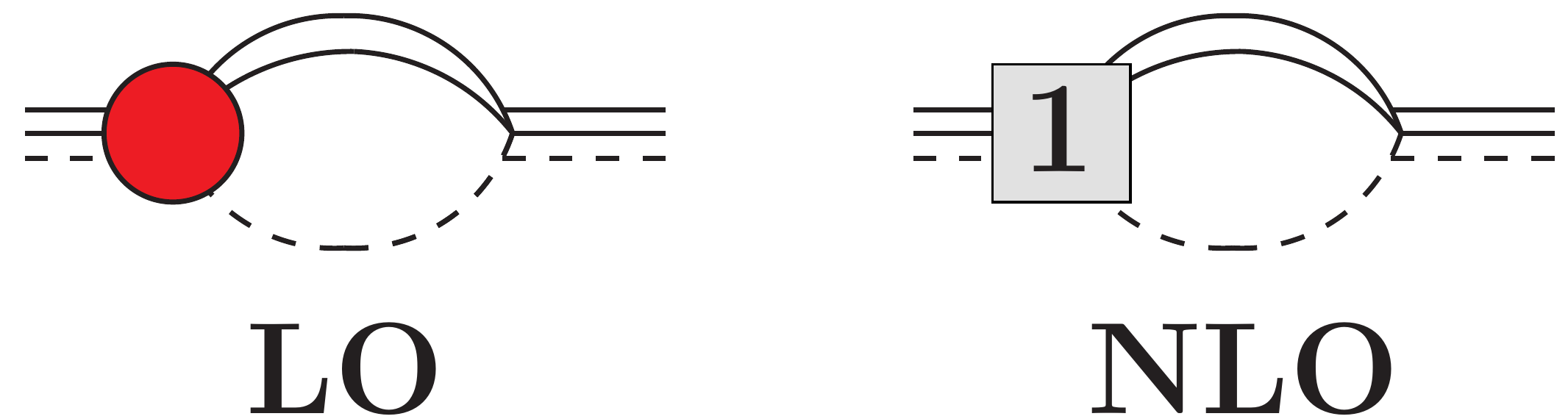}
\caption{\label{fig:Sigmas}Diagrams representing the LO $\Sigma_{0}(E)$ and NLO $\Sigma_{1}(E)$.}
\end{figure}

Finally, the trimer wavefunction renormalization up to NLO is given by
\begin{equation}
Z_{t}=\frac{1}{\Sigma_{0}'(E)}\left[\underbrace{1\vphantom{\frac{\Sigma_{1}'(E)}{\Sigma_{0}'(E)}}}_{\mathrm{LO}}-\underbrace{\frac{\Sigma_{1}'(E)}{\Sigma_{0}'(E)}}_{\mathrm{NLO}}+\cdots\right],
\end{equation}
where $\Sigma_{m}(E)$ are the order-by-order corrections to the trimer self energy defined by~\cite{Vanasse:2015fph}
\begin{equation}
\Sigma_{m}(E)=\frac{m_{n}}{2}\G_{m,t\to d_{0}c}(E,q)\otimes \bar{D}_{nn}^{(0)}(E,q).
\end{equation}
The functions $\Sigma_{0}(E)$ and $\Sigma_{1}(E)$ are given by the diagrams in Fig.~\ref{fig:Sigmas}.
Fitting the energy, $E=B_{cnn}$, to the $cnn$ bound state energy, $B_{cnn}$, yields the values
\begin{equation}
\frac{h_{0}^{2}}{\Omega}=\frac{1}{\Sigma_{0}(E)}\quad,\quad\frac{2h_{1}h_{0}}{\Omega}=-\frac{\Sigma_{1}(E)}{\left(\Sigma_{0}(E)\right)^{2}},
\end{equation}
for the parameters in the three-body Lagrangian~\cite{Vanasse:2015fph}.  However, for the purposes of this calculation the values of three-body forces are not relevant, but only the values of $\Sigma_{0}'(E)$, and $\Sigma_{1}'(E)$ are relevant.  Finally, we define the quantity $Z_{t}^{\mathrm{LO}}$ as
\begin{equation}
Z_{t}^{\mathrm{LO}}=\frac{1}{\Sigma_{0}'(E)}.
\end{equation}

\section{\label{sec:formfactor} Charge and Matter Form Factors}

\begin{figure}[hbt]
\includegraphics[width=110mm]{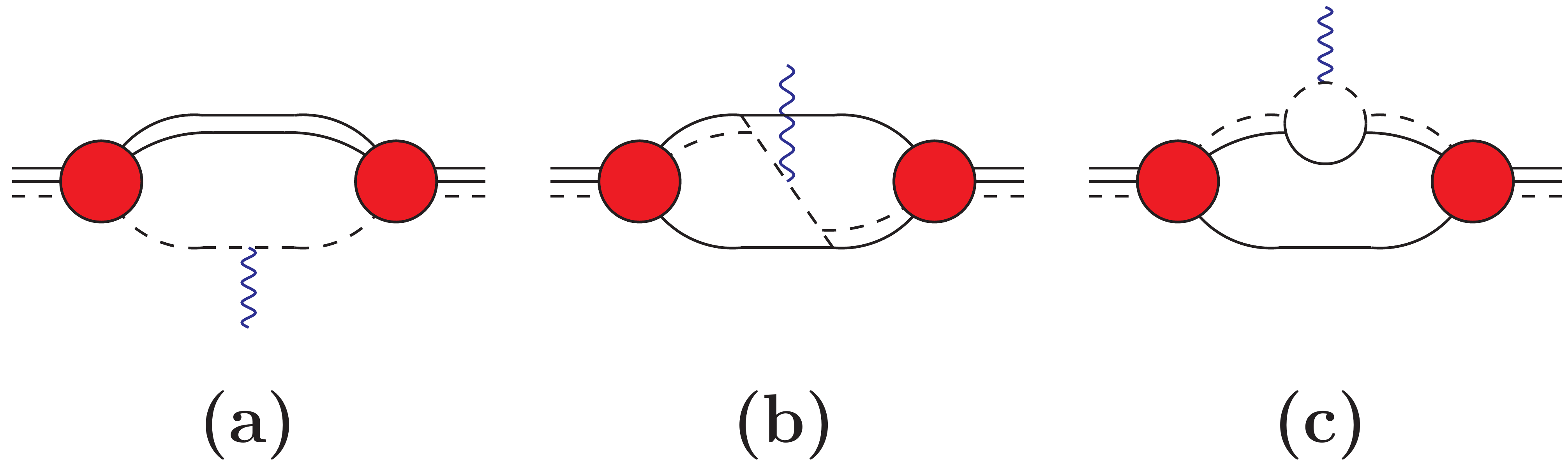}
\caption{\label{fig:HaloLOcharge}Diagrams for the LO charge form factor of two-neutron halo nuclei.  The wavy  blue lines represent minimally coupled $A_{0}$ photons.}
\end{figure}

\begin{figure}[hbt]
\includegraphics[width=110mm]{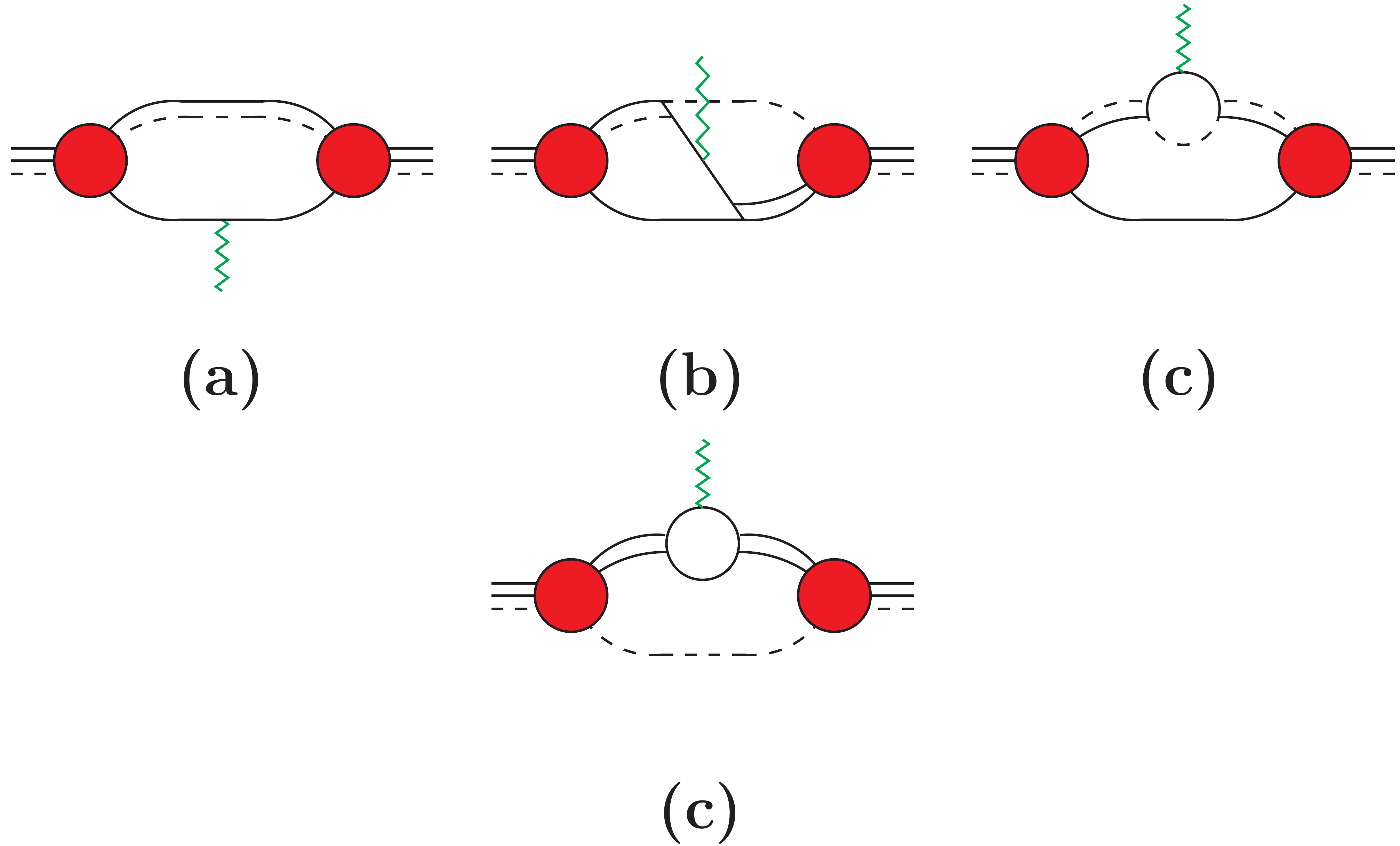}
\caption{\label{fig:HaloLOmatter}Diagrams for the LO neutron form factor of two-neutron halo nuclei.  The green zig-zags represent an external current that only couples to neutrons.  Note there are two (c) type diagrams, one with an intermediate $cn$-dimer and the other with an intermediate $nn$-dimer.}
\end{figure}

The LO charge form factor of two-neutron halo nuclei is given by the sum of diagrams in Fig.~\ref{fig:HaloLOcharge}, where the blue wavy lines represent minimally coupled $A_{0}$ photons that only couple to the charged core.
Meanwhile, the LO neutron form factor of two-neutron halo nuclei is given by the sum of diagrams in Fig.~\ref{fig:HaloLOmatter}, where the green zig-zag is a fictitious current that couples to neutrons with a charge of one.  For the neutron form factor their are two different type (c) diagrams, one for an intermediate $cn$-dimer and the other  for an intermediate $nn$-dimer.  All form factors are calculated in the Breit frame in which the external current only imparts momentum, but no energy.  Form factors only depend on the external current exchange momentum squared, $\vect{Q}^{2}=(\vect{P}-\vect{K})^{2}$, where $\vect{K}$ ($\vect{P}$) is the trimer momentum before (after) the external current.  

The LO diagram (a) contribution for both charge and neutron form factors is given by
\begin{align}
\label{eq:FLOa}
&F_{0}^{(a;X)}(Q^{2})=Z_{t}^{\mathrm{LO}}\left\{\widetilde{\Gb}_{0}^{T}(p)\otimes \boldsymbol{\mathcal{A}}_{0}^{(X)}(p,k,Q)\otimes\widetilde{\Gb}_{0}(k)+2\widetilde{\Gb}_{0}^{T}(p)\otimes \boldsymbol{\mathcal{A}}_{0}^{(X)}(p,Q)+\mathcal{A}_{0}^{(X)}(Q)\right\},
\end{align}
where the superscript $X\!=\!C$ ($X\!=\!n$) for the charge (neutron) form factor.  Functions $\boldsymbol{\mathcal{A}}_{m}^{(X)}(p,k,Q)$, $\boldsymbol{\mathcal{A}}_{m}^{(X)}(p,Q)$, and $\mathcal{A}_{m}^{(X)}(Q)$ are a matrix, vector, and scalar respectively and are defined in Appendix~\ref{app:Func}.  The vector $\widetilde{\Gb}_{m}(p)$ is defined as
\begin{equation}
\widetilde{\Gb}_{m}(p)=\mathbf{D}(E,p)\Gb_{m}(E,p).
\end{equation}
Diagram (b) gives the contribution 
\begin{align}
\label{eq:FLOb}
&F_{0}^{(b;X)}(Q^{2})=Z_{t}^{\mathrm{LO}}\widetilde{\Gb}_{0}^{T}(p)\otimes \boldsymbol{\mathcal{B}}_{0}^{(X)}(p,k,Q)\otimes\widetilde{\Gb}_{0}(k),
\end{align}
to the charge and neutron form factors, where $\boldsymbol{\mathcal{B}}_{0}^{(X)}(p,k,Q)$ is a matrix defined in Appendix~\ref{app:Func}.  The function $\boldsymbol{\mathcal{B}}_{0}^{(X)}(p,k,Q)$ does not receive higher order corrections.  Finally, the contribution from (c) type diagrams to charge and neutron form factors is given by
\begin{align}
\label{eq:FLOc}
&F_{0}^{(c;X)}(Q^{2})=Z_{t}^{\mathrm{LO}}\left\{\widetilde{\Gb}_{0}^{T}(p)\otimes \boldsymbol{\mathcal{C}}_{0}^{(X)}(p,k,Q)\otimes\widetilde{\Gb}_{0}(k)+\boldsymbol{\mathcal{C}}_{0}^{(X)}(k,Q)\otimes\widetilde{\Gb}_{0}(k)+\mathcal{C}_{0}^{(X)}(Q)\right\},
\end{align}
where $\boldsymbol{\mathcal{C}}_{m}^{(X)}(p,k,Q)$ is a matrix, $\boldsymbol{\mathcal{C}}_{m}^{(X)}(k,Q)$ a vector, and $\mathcal{C}_{m}^{(X)}(Q)$ a scalar defined in Appendix~\ref{app:Func}.  Combining the contributions from (a) through (c) type diagrams yields the LO charge and neutron form factors
\begin{equation}
F_{0}^{(X)}(Q^{2})=F_{0}^{(a;X)}(Q^{2})+F_{0}^{(b;X)}(Q^{2})+F_{0}^{(c;X)}(Q^{2}).
\end{equation}

\begin{figure}[hbt]
\includegraphics[width=110mm]{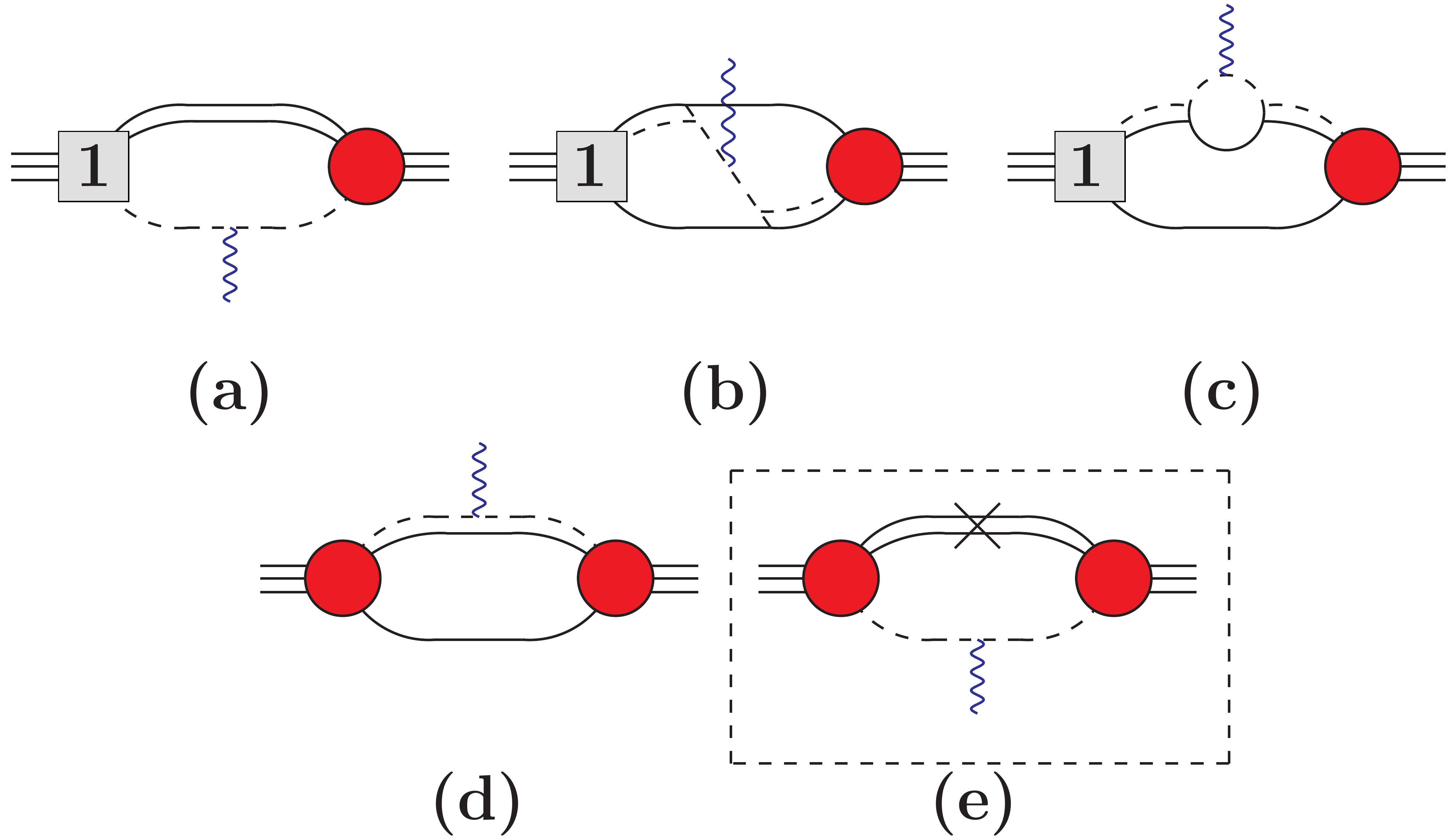}
\caption{\label{fig:HaloNLOcharge}Diagrams for NLO correction to the charge form factor for two-neutron halo nuclei.  The boxed diagram (e) is subtracted to avoid double counting from diagram (a) and its time reversed version, and diagram (d) comes from gauging the $cn$-dimer kinetic term.  Diagrams related by time reversal symmetry are not shown.}
\end{figure}

\begin{figure}[hbt]
\begin{center}
\includegraphics[width=110mm]{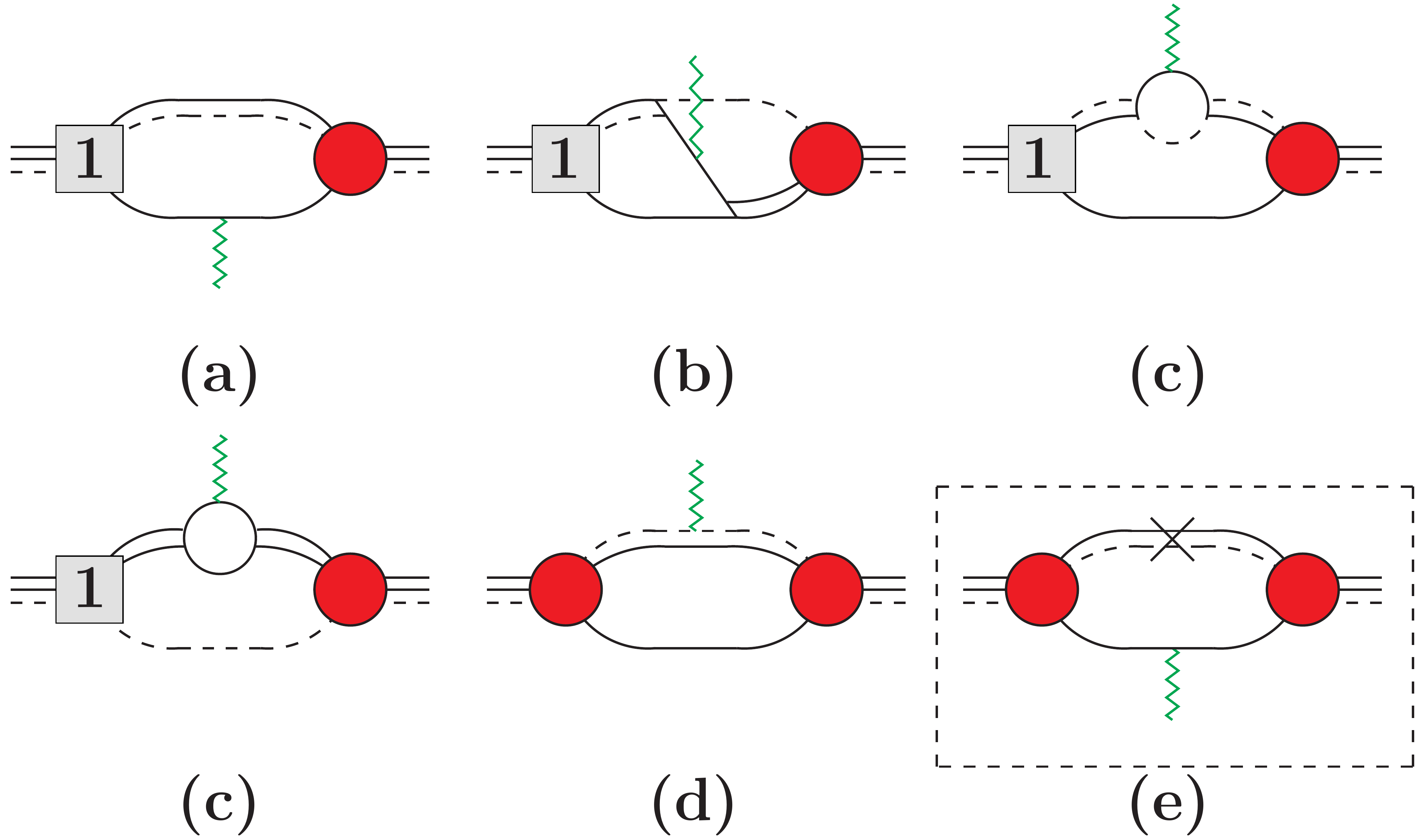}\\
\includegraphics[width=35mm]{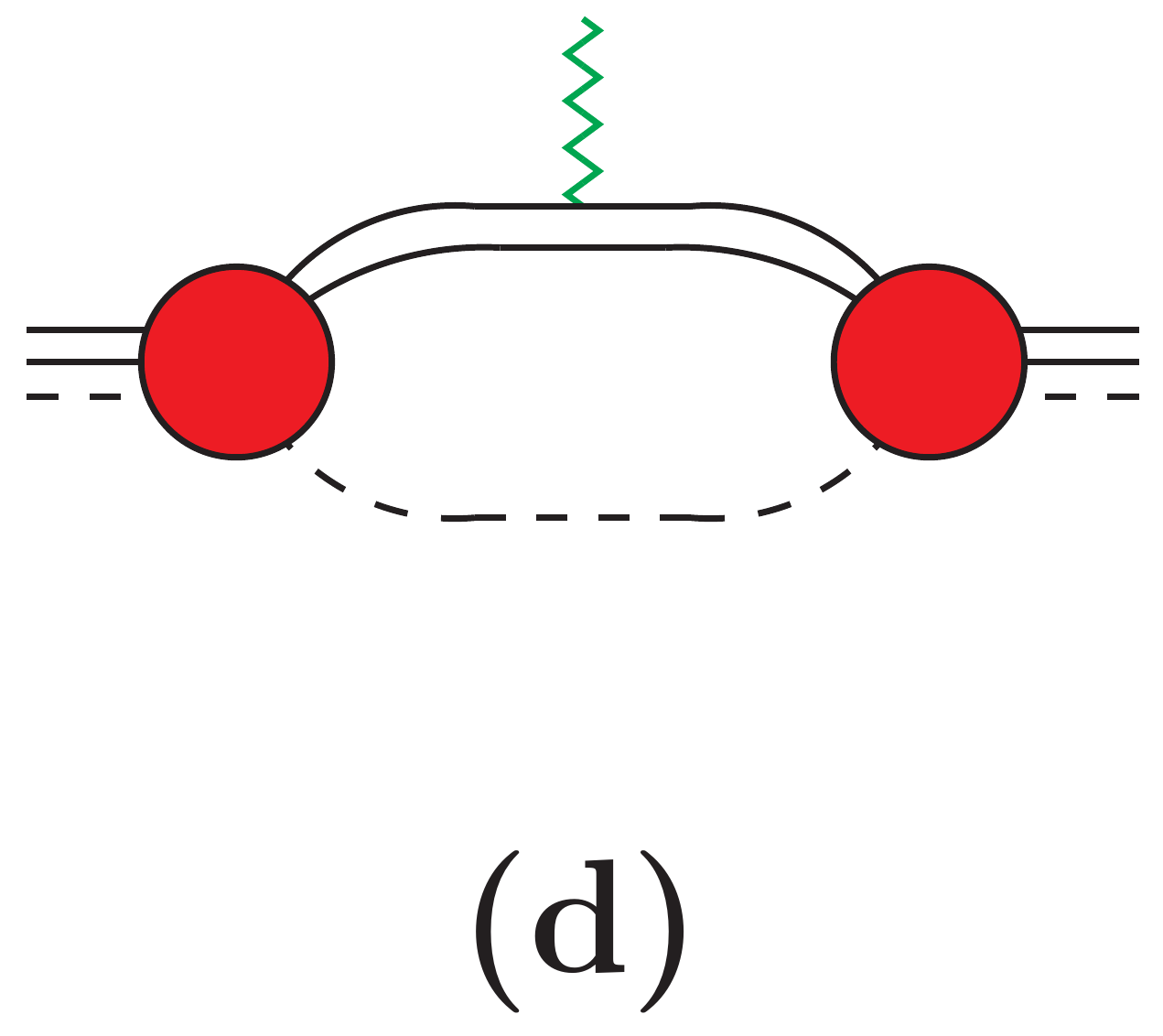}
\end{center}
\caption{\label{fig:HaloNLOmatter}Diagrams for the NLO correction to the neutron form factor for two-neutron halo nuclei.  The boxed diagram (e) is subtracted to avoid double counting.  Note, there are two type (c) and type (d) diagrams.  Diagrams related by time reversal symmetry are not shown}
\end{figure}

The NLO correction to the two-neutron halo nuclei charge form factor is given by the sum of diagrams in Fig.~\ref{fig:HaloNLOcharge}.  Diagram (d) comes from gauging the $cn$-dimer kinetic term in Eq.~(\ref{eq:L2}).
The NLO correction to the two-neutron halo nuclei neutron form factor is given by the sum of diagrams in Fig.~\ref{fig:HaloNLOmatter}, where the coupling for the (d) type diagrams comes from gauging the $nn$- and $cn$-dimer kinetic terms in Eq.~(\ref{eq:L2}).  In Figs.~\ref{fig:HaloNLOcharge} and \ref{fig:HaloNLOmatter}  diagrams related by time reversal symmetry are not shown, and diagram (e) is subtracted to avoid double counting from diagram (a) and its time reversed version.

The NLO correction to the charge and neutron form factor from diagram (a) minus diagram (e) is given by
\begin{align}
\label{eq:NLOatriton}
&F_{1}^{(a;X)}(Q^{2})=Z_{t}^{\mathrm{LO}}\left\{\widetilde{\Gb}_{0}^{T}(p)\otimes \boldsymbol{\mathcal{A}}_{1}^{(X)}(p,k,Q)\otimes\widetilde{\Gb}_{0}(k)+2\widetilde{\Gb}_{1}^{T}(p)\otimes \boldsymbol{\mathcal{A}}_{0}^{(X)}(p,k,Q)\otimes\widetilde{\Gb}_{0}(k)\right.\\\nonumber
&\left.\hspace{5cm}+2\widetilde{\Gb}_{0}^{T}(p)\otimes \boldsymbol{\mathcal{A}}_{1}^{(X)}(p,Q)+ 2\widetilde{\Gb}_{1}^{T}(p)\otimes \boldsymbol{\mathcal{A}}_{0}^{(X)}(p,Q)+\mathcal{A}_{1}^{(X)}(Q)\right\}.
\end{align}
NLO corrections to the charge and neutron form factors from diagram (b) yield
\begin{align}
&F_{1}^{(b;X)}(Q^{2})=Z_{t}^{\mathrm{LO}}\left\{2\widetilde{\Gb}_{1}^{T}(p)\otimes \boldsymbol{\mathcal{B}}_{0}^{(X)}(p,k,Q)\otimes\widetilde{\Gb}_{0}(k)\right\},
\end{align}
and from diagrams (c)
\begin{align}
&F_{1}^{(c;X)}(Q^{2})=Z_{t}^{\mathrm{LO}}\left\{\widetilde{\Gb}_{0}^{T}(p)\otimes \boldsymbol{\mathcal{C}}_{1}^{(X)}(p,k,Q)\otimes\widetilde{\Gb}_{0}(k)+\widetilde{\Gb}_{1}^{T}(p)\otimes \boldsymbol{\mathcal{C}}_{0}^{(X)}(p,k,Q)\otimes\widetilde{\Gb}_{0}(k)\right.\\\nonumber
&\hspace{5cm}+\widetilde{\Gb}_{0}^{T}(p)\otimes \boldsymbol{\mathcal{C}}_{0}^{(X)}(p,k,Q)\otimes\widetilde{\Gb}_{1}(k)+\boldsymbol{\mathcal{C}}_{1}^{(X)}(k,Q)\otimes\widetilde{\Gb}_{0}(k)\\\nonumber
&\left.\hspace{9cm}+\boldsymbol{\mathcal{C}}_{0}^{(X)}(k,Q)\otimes\widetilde{\Gb}_{1}(k)+\mathcal{C}_{1}^{(X)}(Q)\right\}.
\end{align}
At NLO there are new contributions from (d) type diagrams to the charge and neutron form factors which give
\begin{align}
&F_{1}^{(d;X)}(Q^{2})=Z_{t}^{\mathrm{LO}}\left\{\widetilde{\Gb}_{0}^{T}(p)\otimes \boldsymbol{\mathfrak{D}}_{1}^{(X)}(p,k,Q)\otimes\widetilde{\Gb}_{0}(k)+\boldsymbol{\mathfrak{D}}_{1}^{(X)}(k,Q)\otimes\widetilde{\Gb}_{0}(k)+\mathfrak{D}_{1}^{(X)}(Q)\right\},
\end{align}
where $\boldsymbol{\mathfrak{D}}_{1}^{(X)}(p,k,Q)$ is a matrix, $\boldsymbol{\mathfrak{D}}_{1}^{(X)}(k,Q)$ a vector, and $\mathfrak{D}_{1}^{(X)}(Q)$ a scalar defined in Appendix~\ref{app:Func}.  Combining the contribution from diagrams (a) through (d) and multiplying the LO form factor by the NLO trimer wavefunction renormalization gives the NLO correction to the charge and neutron form factors
\begin{equation}
F_{1}^{(X)}(Q^{2})=\left(F_{1}^{(a;X)}(Q^{2})+F_{1}^{(b;X)}(Q^{2})+F_{1}^{(c;X)}(Q^{2})+F_{1}^{(d;X)}(Q^{2})\right)-\frac{\Sigma_{1}'(E)}{\Sigma_{0}'(E)}F_{0}^{(X)}(Q^{2}).
\end{equation}

\section{\label{sec:observable} Observables}

Expanding the LO two-neutron halo nuclei charge form factor as a function of $Q^{2}$ yields
\begin{equation}
F_{0}^{(C)}(Q^{2})=1-\frac{1}{6}\left<r_{C}^{2}\right>_{0}Q^{2}+\cdots
\end{equation}
where $\left<r_{C}^{2}\right>_{0}$ is the LO point charge radius squared of the $cnn$ system.  The LO neutron form factor expanded in powers of $Q^{2}$ yields
\begin{equation}
F_{0}^{(n)}(Q^{2})=2\left(1-\frac{1}{6}\left< r_{n}^{2}\right>_{0}Q^{2}+\cdots\right)
\end{equation}
where $\left<r_{n}^{2}\right>_{0}$ is the LO neutron radius of the $cnn$ system.  Expanding in powers of $Q^{2}$ the NLO correction to the charge form factor is given by
\begin{equation}
F_{1}^{(C)}(Q^{2})=-\frac{1}{6}\left<r_{C}^{2}\right>_{1}Q^{2}+\cdots,
\end{equation}
and the NLO correction to the neutron form factor by
\begin{equation}
F_{1}^{(n)}(Q^{2})=2\left(-\frac{1}{6}\left<r_{n}^{2}\right>_{1}Q^{2}+\cdots.\right),
\end{equation}
where $\left<r_{n}^{2}\right>_{1}$ ($\left<r_{C}^{2}\right>_{1}$)  is the NLO correction to the neutron (point charge) radius of the $cnn$ system.  Due to gauge invariance the NLO correction to the form factors are zero at $Q^{2}=0$ and this is observed numerically to at least seven digits.  Likewise it is observed that $F_{0}^{(C)}(0)=1$ and $F_{0}^{(n)}(0)=2$ to at least seven digits.

The point charge radius squared of the $cnn$ system is related to its physical charge radius squared, $\left<r_{C}^{2}\right>^{cnn}$, by
\begin{equation}
\left<r_{C}^{2}\right>^{cnn}=\left<r_{C}^{2}\right>+\left<r_{C}^{2}\right>^{c}+\frac{2}{Z}\left<r_{C}^{2}\right>^{n},
\end{equation}
where $\left<r_{C}^{2}\right>^{c}$ is the charge radius squared of the core, $Z$ the number of protons in the core, and $\left<r_{C}^{2}\right>^{n}=-0.115(4)$~fm$^{2}$~\cite{Agashe:2014kda} is the charge radius squared of the neutron.  In isotope shift experiments using laser spectroscopy the value of $\left<r_{C}^{2}\right>$ is directly accessible if the relatively small contribution from the neutron charge radius squared is ignored~\cite{Sanchez:2006zz}.  The point matter radius of the $cnn$ system is obtained from the charge and neutron radius via
%
%\begin{equation}
%\sqrt{\left<r_{M}^{2}\right>}=\sqrt{\frac{1}{3}\left(\left<r_{C}^{2}\right>+2\left<r_{n}^{2}\right>\right)}
%\end{equation}
%
\begin{equation}
\sqrt{\left<r_{M}^{2}\right>}=\sqrt{\frac{1}{2\!+\!A}\left((A\left<r_{C}^{2}\right>+2\left<r_{n}^{2}\right>\right)},
\end{equation}
and the physical matter radius squared, $\left<r_{M}^{2}\right>^{cnn}$, of the $cnn$ system is related to the point matter radius squared by
\begin{equation}
\label{eq:matterR}
\left<r_{M}^{2}\right>^{cnn}=\left<r_{M}^{2}\right>+\frac{A}{2\!+\!A}\left<r_{M}^{2}\right>^{c}+\frac{2}{2\!+\!A}\left<r_{M}^{2}\right>^{n},
\end{equation}
where $\left<r_{M}^{2}\right>^{c}$ is the matter radius squared of the core and $\left<r_{M}^{2}\right>^{n}$ is the matter radius squared of the neutron. The small contribution from the neutron is ignored.

\begin{figure}[hbt]
\includegraphics[width=110mm]{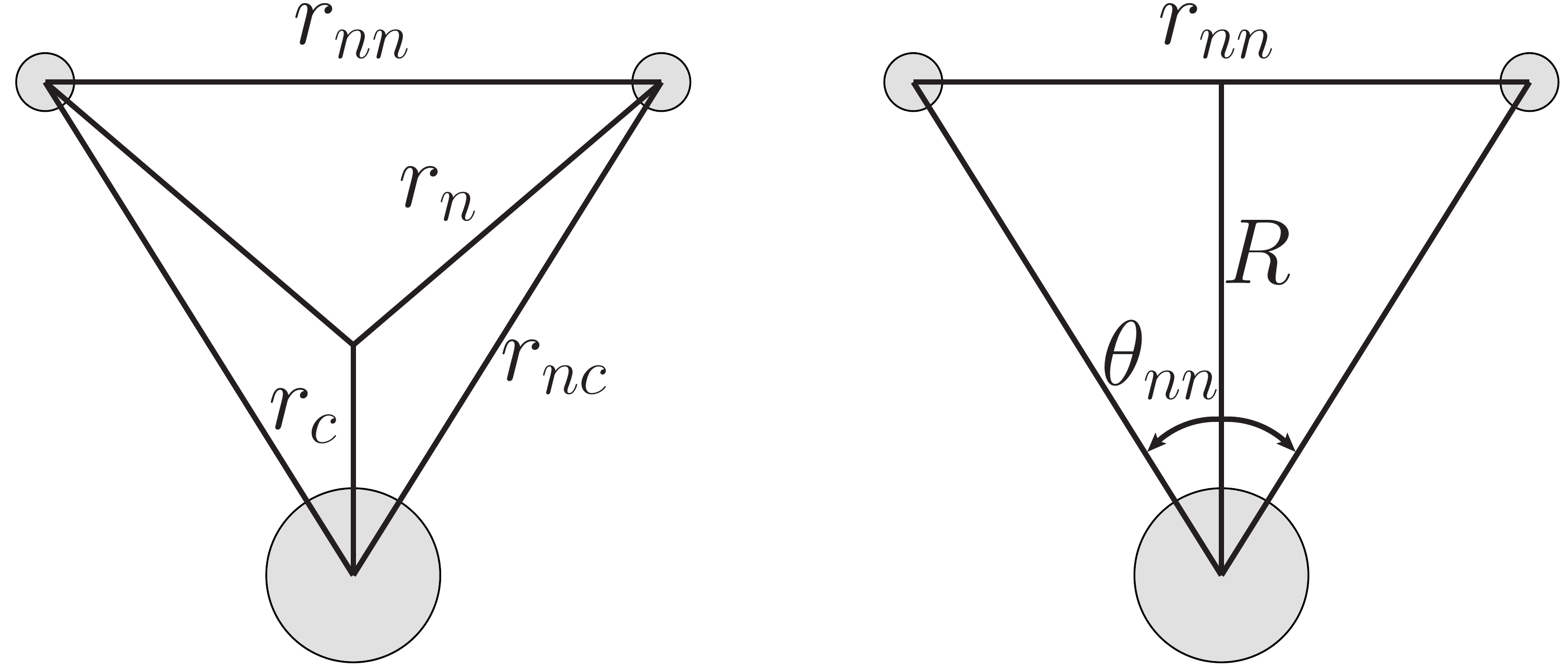}
\caption{Geometric representation of two-neutron halo nucleus.  The value $r_{c}$ is the charge radius and $r_{n}$ the neutron radius, which both extend from from the c.m. of the two-neutron halo nucleus to the core (large circle) and neutron (small circle) respectively.\label{fig:HaloRadii}}
\end{figure}

Two-neutron halo nuclei can be understood geometrically as in Fig.~\ref{fig:HaloRadii}. 
The large circle represents the core and the smaller circles the valence neutrons.  $r_{c}$ is the point charge radius that extends from the center of mass (c.m.) of the $cnn$ system to the core, and $r_{n}$ is the neutron radius.  Writing all other geometrical quantities in Fig.~\ref{fig:HaloRadii} in terms of $r_{c}$, $r_{n}$, and $A$ gives
\begin{equation}
R=\frac{2+A}{2}r_{c}
\end{equation}
for the average distance between the core and the $nn$ center of mass,
\begin{equation}
r_{nn}=2\sqrt{r_{n}^{2}-\left(\frac{A}{2}r_{c}\right)^{2}}
\end{equation}
for the average inter-neutron separation,
\begin{equation}
r_{nc}=\sqrt{r_{n}^{2}+\frac{2+A}{2}r_{c}^{2}}
\end{equation}
for the average core neutron separation, and
\begin{equation}
\theta_{nn}=2\arctan\left(\frac{r_{nn}}{(2+A)r_{c}}\right)
\end{equation}
for the neutron opening angle.  These geometrical quantities prove useful as they are more accessible in certain experiments and also have widespread adoption in the literature.  

\section{\label{sec:results} Results}

The LO calculation of the neutron and charge form factors only requires four two-body inputs and one piece of three-body data.  In the $nn$ channel there is the $nn$ virtual bound state energy, $\gamma_{nn}=-9.87$~MeV, and the neutron mass.  While in the $cn$ channel there is the core mass and $cn$ ``binding energy", $B_{cn}$, given in Table~\ref{tab:parameters} for halo nuclei considered in this work.  For unbound $cn$ systems $|B_{cn}|$ is given by the real part of the lowest lying resonance for $cn$ scattering and $B_{cn}$ is negative.  Using the $cnn$ system binding energy given in Table~\ref{tab:parameters} the three-body force is fixed at LO and NLO.  In addition the quantum numbers of the core, $cn$ system, and bound $cnn$ system are shown in Table~\ref{tab:parameters}.  The Lithium system does not have a spin zero core as assumed in our formalism.  However, since the core is much heavier than the neutrons the static limit in which the core spin is unchanged can be used to approximate the core as spin zero.  Other two-neutron halo nuclei such as ${}^6$He and $^{17}$B are not considered here as they are dominated by two-body $P$-wave interactions and will be dealt with in future work.  At NLO only the two-body $nn$ effective range $\rho_{nn}$ and $cn$ effective range $\rho_{cn}$ are needed.

\begin{table}
\begin{center}
\begin{tabular}{|c|c|c|c|c|}
\hline
Nucleus & $m_{C}$ [MeV] & $B_{cn}$ [MeV] & $B_{cnn}$ [MeV] & $J_{c}^{\pi}:J_{cn}^{\pi}:J_{cnn}^{\pi}$\\\hline
$^{9-11}$Li & 8408 & -0.026(13) & 0.3693(6) & $3/2^{-}$:$(2^{-},1^{-})$:$3/2^{-}$\\
$^{12-14}$Be & 11203 & -0.510 & 1.27(13) & $0^{+}$:($1/2^{-}$):$0^{+}$ \\
$^{20-22}$C & 18667 & -0.01(47)~\cite{1674-1137-36-12-003} & 0.11(6)~\cite{1674-1137-36-12-003} & $0^{+}$:$(1/2^{+})$:$0^{+}$\\\hline
\end{tabular}
\end{center}
\caption{\label{tab:parameters}The mass $m_{c}$ and  quantum numbers, $J^{\pi}_{c}$, of the $^{9}$Li, $^{12}$Be, and $^{20}$C core are given.  Binding energies, $B_{cn}$ and quantum numbers, $J^{\pi}_{cn}$, of the $^{10}$Li, $^{13}$Be, and $^{21}$C resonances are given and the bound state energies, $B_{cnn}$ and quantum numbers, $J^{\pi}_{cnn}$, of the halo nuclei $^{11}$Li, $^{14}$Be, and $^{22}$C.  The quantum number $J$ is the total angular momentum and $\pi$ the parity.  All numbers without a reference come from Ref.~\cite{NuDat}.}
\end{table}

There are three sources of error in the caluclation of the form factors, (i) numerical error, (ii) error from two- and three-body parameters, and (iii) error from the halo-EFT expansion.  Numerical error is negligible compared to the other sources of error and will henceforth be disregarded.  The predominant source of error for $^{11}$Li and $^{14}$Be comes from the halo-EFT expansion, which is estimated in Table~\ref{tab:error} for each halo nucleus, where ratios of $M_{hi}$ to $M_{low}$ parameters are taken.  The scales for $M_{hi}$ are the core's first excited state energy, $E_{C}^{*}$, and the one neutron separation energy, $B_{c-n}$, for the core, both shown in Table~\ref{tab:error}.  These scales signal the breakdown of halo-EFT since the core can no longer be a fundamental degree of freedom at these energies.  The scales for $M_{low}$ are given by $B_{cn}$ and $B_{cnn}$.  Taking the most conservative error estimate we find the error of the EFT expansion is 37\% for $^{11}$Li, 78\% for $^{14}$Be, and 26\% for $^{22}$C.  However, the error for $^{22}$C is dominated by the uncertainty in $B_{cn}$ and $B_{cnn}$.  Since within the error of $B_{cn}$ it can equal $B_{cnn}$ the charge and matter radius of $^{22}$C diverge, and it can only be bounded from below.

\begin{table}
\begin{center}
\begin{tabular}{|c|cc|cccc|}
\hline
Nucleus& $E_{C}^{*}$ [MeV] & $B_{c-n}$ [MeV] & $\sqrt{|B_{cn}/E_{C}^{*}|}$ & $\sqrt{B_{cnn}/E_{C}^{*}}$ & $\sqrt{|B_{cn}/B_{c-n}|}$ & $\sqrt{|B_{cnn}/B_{c-n}|}$\\\hline
$^{9-11}$Li& 2.69 & 4.06 & 0.098 & \textbf{0.370} & 0.078 & 0.301  \\
$^{12-14}$Be& 2.10 & 3.17 & 0.493 & \textbf{0.78} & 0.40 & 0.63\\
$^{20-22}$C& \,\,\,\,\,\, 1.59~\cite{Stanoiu:2008zz} & 2.9(3) & 0.09 & \textbf{0.26} & 0.07 & 0.19 \\\hline
\end{tabular}
\caption{\label{tab:error}The first excited state energy $E_{c}^{*}$ and the one neutron separation energy $B_{c-n}$ of the $^{9}$Li, $^{12}$Be, and $^{20}$C cores are given.  Using these values the ratios $\sqrt{|B_{cn}/E_{C}^{*}|}$, $\sqrt{|B_{cnn}/E_{c}^{*}|}$, $\sqrt{|B_{cn}/B_{c-n}|}$, and $\sqrt{|B_{cn}/B_{c-n}|}$ are calculated to estimate the error of the halo-EFT expansion.  All numbers for $E_{C}^{*}$ and $B_{c-n}$ without a reference come from Ref.~\cite{NuDat}.}
\end{center}
\end{table}
 
The LO and NLO predictions for the point charge radius, point matter radius, and existing experimental determinations are shown in Table~\ref{tab:RadiiResults} for each halo nucleus.
\begin{table}
\begin{center}
\begin{tabular}{|c|c|c|c|c|P{2.7cm}|P{3.0cm}|}
\hline
Nucleus & $\left< r_{C}^{2}\right>_{0}$~fm$^{2}$ & $\left<r_{M}^{2}\right>_{0}$~fm$^{2}$ & $\left< r_{C}^{2}\right>_{0+1}$~fm$^{2}$ & $\left<r_{M}^{2}\right>_{0+1}$~fm$^{2}$ & $\left<r_{C}^{2}\right>$-Exp.~fm$^{2}$ & $\left<r_{M}^{2}\right>$-Exp.~fm$^{2}$  \\\hline
$^{11}$Li & $0.744(275)$ & $5.76\pm2.13$ & $0.774(106)$ & $6.05\pm0.83$  & 1.171(120)~\cite{Sanchez:2006zz}\newline 1.104(85)~\cite{Puchalski:2006zz} \newline 0.82(11)~\cite{Esbensen:2007ir,Nakamura:2006zz} & $5.34\pm0.15$~\cite{Ozawa:2001hb}\\\hline
$^{14}$Be & $0.126(98)$ & $1.23\pm0.96$ & $0.134(81)$ & $1.34\pm0.81$  & --- & $4.24\pm2.42$~\cite{Ozawa:2001hb} $2.90\pm2.25$~\cite{Ozawa:2001hb} \\\hline
$^{22}$C & $0.520^{+\infty}_{-0.274}$ & $9.00^{+\infty}_{-5.01}$ & $0.530^{+\infty}_{-0.283}$ & $9.22^{+\infty}_{-5.16}$ &--- & $21.1\pm9.7$~\cite{Ozawa:2000gx,Tanaka:2010zza} \newline $3.77\pm0.61$~\cite{Ozawa:2000gx,Togano2016} \\\hline
\end{tabular}
\caption{\label{tab:RadiiResults}LO and NLO halo-EFT predicitions for charge and mater radii of two-neutron halo nuclei.  Included are existing experimental results.  The NLO results use the naturalness estimate $\rho_{cn}\sim 1/m_{\pi}\sim1.4$~fm for the NLO prediction, where $\rho_{cn}$ is the effective range for $cn$ scattering.}
\end{center}
\end{table}
All results are calculated at a cutoff of $\Lambda=20000$~MeV, which is sufficient to ensure convergence with respect to $\Lambda$.  The radii are extracted from the form factors by performing a linear fit with respect to $Q^{2}$ over the range $Q^{2}=0-0.5$~MeV$^{2}$.  At NLO the $cn$ effective range, $\rho_{cn}$, is estimated using naturalness assumptions giving the value $\rho_{cn}\sim 1/m_{\pi}= 1.4$~fm.  Atomic spectroscopy gives the experimental value of $1.171(120)$~fm$^{2}$ for the $^{11}$Li point charge radius squared~\cite{Sanchez:2006zz}.  This value was later revised in Ref.~\cite{Puchalski:2006zz} by adding finite mass corrections giving a value of 1.104(85)~fm$^{2}$.  Using the cluster sum rule~\cite{Bertsch:1991zz} the experimental electric dipole response of $^{11}$Li~\cite{Nakamura:2006zz} can be related to its point charge radius squared yielding a value of $0.82(11)$~fm$^{2}$~\cite{Esbensen:2007ir}.  Our LO prediction for the $^{11}$Li point charge radius of $0.744(275)$~fm$^{2}$ squared agrees with the smaller experimental value within errors, and NLO corrections, assuming a natural value for $\rho_{cn}$, give $0.774(106)$~fm$^{2}$ again agreeing with the smaller experimental number within errors.  The difference between these experimental values is often attributed to polarization effects of the core~\cite{Esbensen:2007ir}, which occur at orders beyond NLO in halo-EFT.  In addition realistic values of $\rho_{cn}$ may alleviate some of the disagreement with atomic spectroscopy measurements.  Range corrections are found to be an important contribution for the triton charge radius in \EFT~\cite{Vanasse:2015fph}.

The point charge radius squared of $^{14}$Be is $0.126(98)$~fm$^{2}$ at LO and $0.134(81)$~fm$^{2}$ at NLO, while for $^{22}$C we find $0.520^{+\infty}_{-0.274}$~fm$^{2}$ at LO and $0.530^{+\infty}_{-0.283}$~fm$^{2}$ at NLO.  The range for $^{22}$C comes from varying $B_{cn}$ and $B_{cnn}$ within their experimental errors.  Unfortunately, no experimental determination of the charge radius currently exists for $^{14}$Be or $^{22}$C.  Our results for the point charge radius of $^{11}$Li, $^{14}$Be, and $^{22}$C disagree with those of Hagen et \emph{al}.~\cite{Hagen:2013xga}.  However, if we change the coefficient of a single $Q^{2}$ term in diagram (a) of Fig.~\ref{fig:HaloLOcharge} then we reproduce the charge radii of Hagen et \emph{al}.~and moreover agree with their expressions for the diagrams in Fig.~\ref{fig:HaloLOcharge}.  For further details of this difference see Appendix~\ref{app:Func}.

The point matter radius squared of $^{11}$Li is $5.76\pm2.13$~fm$^{2}$ at LO and $6.05\pm0.83$~fm$^{2}$ at NLO.  This agrees well with the experimental number for the $^{11}$Li point matter radius of $5.34\pm0.15$~fm$^{2}$, which is given by the matter radius of $2.32\pm0.02$~fm~\cite{Ozawa:2001hb} for $^{9}$Li, the matter radius of $3.12\pm0.02$~fm~\cite{Ozawa:2001hb} for $^{11}$Li, and the use of Eq.~(\ref{eq:matterR}).  In Eq.~(\ref{eq:matterR}) the unknown matter radius of the neutron is ignored since it is suppressed by a factor of $2/(2\!+\!A)$.  For $^{14}$Be we find a LO point matter radius squared of $1.23\pm0.96$~fm$^{2}$ and NLO value of $1.34\pm0.81$~fm$^{2}$, which agrees with the experimental results of $4.24\pm2.42$~fm$^{2}$ $2.90\pm2.25$~fm$^{2}$ within the large theoretical and experimental uncertainty.  $^{22}$C has a LO point matter radius squared of $9.00^{+\infty}_{-5.01}$~fm$^{2}$ and a NLO matter radius squared of $9.22^{+\infty}_{-5.16}$~fm$^{2}$.  The ranges for $^{22}$C, due to varying $B_{cn}$ and $B_{cnn}$ within their experimental errors, overlap with the experimental results of $21.1\pm9.7$~fm$^{2}$ and $3.77\pm0.61$~fm$^{2}$.  In order to find the experimental point matter radius of $^{14}$Be and $^{22}$C we used Eq.~(\ref{eq:matterR}), the value $2.59\pm0.06$~fm~\cite{Ozawa:2001hb} for the $^{12}$Be matter radius, $3.16\pm0.38$~fm~\cite{Ozawa:2001hb} and $2.94\pm0.38$~fm~\cite{Ozawa:2001hb} for the $^{14}$Be matter radius, $2.98\pm0.05$~fm~\cite{Ozawa:2001hb} for the $^{20}$C~\cite{Ozawa:2000gx} matter radius, and $5.4\pm0.9$~fm~\cite{Tanaka:2010zza} and $3.44\pm0.08$~fm~\cite{Togano2016} for the $^{22}$C matter radius.  The smaller values for the $^{14}$Be and $^{22}$C matter radii give the smaller experimental values for their respective point matter radii in Table~\ref{tab:RadiiResults}.

In addition to the point matter and point charge radii the LO values for the average inter-neutron distance, $r_{nn}$, and neutron opening angle, $\theta_{nn}$, shown in Fig.~\ref{fig:HaloRadii}, are given in Table~\ref{tab:rnnAndthetann}.  Experimental results of $r_{nn}$ for $^{11}$Li and $^{14}$Be from Marqu\'{e}s et \emph{al.}~\cite{Marques:2000mqq,Marques:2001pe} using two-neutron interferometry agree within errors with our predictions.  The neutron opening angles of Bertulani et \emph{al.}~\cite{Bertulani:2007rm} are determined using $r_{nn}$ from Marqu\'{e}s et \emph{al.}~\cite{Marques:2000mqq,Marques:2001pe}, atomic spectroscopy data on $^{11}$Li for $R$~\cite{Sanchez:2006zz}, which gives an angle of $58^{\circ+10}_{-14}$, and dipole response data for $^{11}$Li with the cluster sum rule for $R$, which gives an angle of $66^{\circ+22}_{-18}$~\cite{Nakamura:2006zz}.  For $^{14}$Be Bertulani et \emph{al.}~\cite{Bertulani:2007rm} used a model calculation for $R$.  The neutron opening angles of Hagino et \emph{al.}~\cite{Hagino:2007rn} use the dipole response data for $^{11}$Li with a model to extract $R$, the matter radius of $^{11}$Li~\cite{Ozawa:2001hb} to get the neutron opening angle $56.2^{\circ+17.8}_{21.3}$, and the data of Marqu\'{e}s et \emph{al.}~\cite{Marques:2000mqq,Marques:2001pe} to get $r_{n}$, which gives a value of $65.2^{\circ+11.4}_{-13.0}$.  These values agree within errors with our calculated results.  However, the error bars are quite large.  Also shown in Table~\ref{tab:rnnAndthetann} are LO halo-EFT predictions from Canham and Hammer~\cite{Canham:2008jd} for $^{11}$Li and $^{14}$Be.  For $^{11}$Li they find the neutron opening angles $77^{\circ+8}_{-9}$ and $68^{\circ+31}_{-25}$ using different values of $B_{cn}$ and for $^{14}$Be find $72^{\circ+16}_{-13}$. Their values for $\theta_{nn}$ and $r_{nn}$ differ from ours in part due to different choices for the values of $B_{cn}$ and $B_{cnn}$, however, within errors they agree with our results.  These calculations were extended to NLO in Ref.~\cite{Canham:2009xg} by resumming range corrections to all orders and using naturalness assumptions for $\rho_{cn}$.  This differs from this work in which range corrections are added perturbatively.  NLO values for $\theta_{nn}$ and $r_{nn}$ are not shown because $\theta_{nn}$ barely changes and $r_{nn}$ only slightly.

\begin{table}
\begin{center}
\begin{tabular}{|c|c|c|P{2.8cm}|P{3.9 cm}|}
%\begin{tabular}{>{\centering\arraybackslash}p{1cm}}
\hline
Nucleus & $r_{nn}$~fm & $\theta_{nn}$ deg. & $r_{nn}$~fm & $\theta_{nn}$ deg.\\\hline
$^{11}$Li & 7.30$^{+1.24}_{-1.51}$ & 75.4$^{+43.1}_{-36.3}$ & $6.6\pm1.5$~\cite{Marques:2000mqq,Marques:2001pe}
\newline $8.7\pm 0.7$~\cite{Canham:2008jd} \newline $6.8\pm 1.8$~\cite{Canham:2008jd}  & 58$^{+10}_{-14}$, 66$^{+22}_{-18}$~\cite{Bertulani:2007rm}\newline 56.2$^{+17.8}_{-21.3}$, 65.2$^{+11.4}_{-13.0}$~\cite{Hagino:2007rn} \newline $77^{+8}_{-9}$, $68^{+31}_{-25}$~\cite{Canham:2008jd} \\\hline
$^{14}$Be & 3.66$^{+1.22}_{-1.94}$ & 73.1$^{+88.0}_{-62.6}$ & $5.4\pm1.0$~\cite{Marques:2000mqq,Marques:2001pe}\newline $4.1\pm 0.5$~\cite{Canham:2008jd} & 64$^{+9}_{-10}$~\cite{Bertulani:2007rm} \newline $72^{+16}_{-13}$~\cite{Canham:2008jd}\\\hline
$^{22}$C & 13.0$^{+\infty}_{-4.72}$ & 78.8$^{+101.2}_{-27.3}$ & --- & --- \\\hline
\end{tabular}
\caption{\label{tab:rnnAndthetann}Values of $r_{n}$ and $\theta_{nn}$ for halo nuclei.  The error for $r_{nn}$ and $\theta_{nn}$ on $^{22}$C is due to varying the value for $B_{cn}$ and $B_{cnn}$ within their errors. }
\end{center}
\end{table}

Table~\ref{tab:NLOcorrections} gives the NLO corrections to the charge and matter point radii from the $nn$ and $cn$ effective range corrections separately.  The $nn$ NLO range corrections use the physical values for the $nn$ effective range correction, whereas for the $cn$ effective range
\begin{equation}
\frac{Z_{cn}-1}{2\gamma_{cn}}=1\, \mathrm{MeV}^{-1}.
\end{equation}
If future experiments determine the $cn$ effective range $\rho_{cn}$ then the value $\frac{Z_{cn}-1}{2\gamma_{cn}}$ can be calculated and multiply the results in Table~\ref{tab:NLOcorrections} to get the physical $cn$ NLO range corrections.

\begin{table}
\begin{center}
\begin{tabular}{|c|c|c|c|c|}
\hline
Nucleus & $nn$: $\left< r_{C}^{2}\right>_{1}$~fm$^{2}$ &$nn$: $\left<r_{M}^{2}\right>_{1}$~fm$^{2}$ & $cn$: $\left<r_{C}^{2}\right>_{1}$~fm$^{2}$ & $cn$: $\left<r_{M}^{2}\right>_{1}$~fm$^{2}$  \\\hline
$^{11}$Li &  $3.71\times 10^{-3}$ & 0.117  & 7.66 & 49.2\\\hline
$^{14}$Be & $2.05\times10^{-3}$ & $5.94\times10^{-2}$  & 1.99 & 15.1\\\hline
$^{22}$C & -(${4.18}^{+\infty}_{-0.587}\times 10^{-4}$) & $6.75^{+\infty}_{-1.02}\times10^{-2}$  &$3.02^{+\infty}_{-2.38}$ & $43.8^{+\infty}_{-40.5}$\\\hline
\end{tabular}
\caption{\label{tab:NLOcorrections}NLO halo-EFT corrections for charge and mater radii of two-neutron halo nuclei.  The $nn$: results come from setting $nn$ effective range corrections to their physical values and setting $cn$ effective range corrections to zero, while the $cn$: results come from setting $nn$ effective range corrections to zero and setting the quantity $(Z_{cn}-1)/(2\gamma_{cn})=1~\mathrm{MeV}^{-1}$.}
\end{center}
\end{table}

Finally, in the unitary and equal mass limit their exists an analytical result~\cite{Braaten:2004rn} for the point charge and point matter radius squared which states
\begin{equation}
\label{eq:unitary}
mE\left<r^{2}\right>=\frac{1+s_{0}^{2}}{9}=0.224...,
\end{equation}
where $m$ is the mass of the particles, $E$ the three-body binding energy, $\left<r^{2}\right>$ the point charge or point matter radius squared, and $s_{0}\approx1.00624$~\cite{Danilov:1963} is a universal number from the asymptotic solution of the three-boson problem with short range interactions. Taking the equal mass and unitary limit in our code we find the number 0.224 for the combination of parameters in Eq.~(\ref{eq:unitary}).  Note, that any technique that claims to be able to calculate the zero-range limit exactly must obtain this result within numerical accuracy.  This number should serve as an essential benchmark for any technique claiming to calculate three-body systems in the zero range approximation.

\section{\label{sec:conclsuion} Conclusion}

Using halo-EFT to NLO we have calculated the charge and neutron form factors for the two-neutron halo nuclei $^{11}$Li, $^{14}$Be, and $^{22}$C.  From the form factors we extracted the point charge and point matter radii to NLO as well as the inter-neutron separation and neutron opening angle to LO.  NLO results were obtained using a naturalness assumption for the $cn$ effective range, $\rho_{cn}\sim1/m_{\pi}=1.4$~fm. At LO and NLO agreement was found between the predicted matter radii and experimental extractions.  However, this is partly due to the large error bars in both experiment and theory.  Further work will be needed in both theory and experiment to further reduce these error bars. The charge radius of $^{11}$Li was found to agree with the experimental extraction from the electric dipole response function of $^{11}$Li, but found to slightly under-predict the charge radius from laser spectroscopy.  Charge radii for $^{14}$Be and $^{22}$C were also given for which there are no current experimental determinations.  Future experiments measuring the charge form factors of halo nuclei are planned for the electron-ion scattering experiment (ELISe) at the Internationl Facility for  Antiproton and Ion Research (FAIR)~\cite{Antonov:2011zza}.

The inter-neutron separation and neutron opening angle were also calculated and compared with experimental extractions.  Again agreement was found with ``experimental" values, but this is in part due to large error bars.  Only LO values are shown for these numbers as the neutron opening angle barely changes at NLO and the inter-neutron separation only slightly.  Finally, the NLO corrections to the point charge and point matter radii from the $nn$ effective range $\rho_{nn}$ and the $cn$ effective range $\rho_{cn}$ were calculated separately, such that the point charge and point matter radii can be easily calculated to NLO once $\rho_{cn}$ is measured.

The point charge and point matter radii were also calculated in the unitary equal mass limit and shown to agree with the analytical prediction of Ref.~\cite{Braaten:2004rn}.  However, our point charge radii disagree with those of Hagen et \emph{al.}~\cite{Hagen:2013xga}.  Comparing our functions for the LO charge form factor with those of Hagen et \emph{al.} we find a minor discrepancy given in detail in Appendix~\ref{app:Func}.  Using the incorrect function from Hagen et \emph{al.} we reproduce the point charge radii given in their paper, but fail to reproduce the correct value in the unitary and equal mass limit.

In order to have more realistic predictions at NLO the parameter $\rho_{cn}$ must be known.  One possible way to measure $\rho_{cn}$ for $n+ ^{9}$Li is through the breakup process $d(^{9}\mathrm{Li})\to np(^{9}\mathrm{Li})$.  Certain kinematical regimes of the three-body breakup spectrum should be especially sensitive to the $n ^{9}$Li interaction.  A halo-EFT calculation of this process is complicated by the binding energy of the deuteron, 2.22~MeV, being only slightly smaller than the first excited state energy of $^{9}$Li, 2.69~MeV.  The ratio of these two quantities makes for a poor expansion and would likely require that the first excited state of $^{9}$Li be added as a new degree of freedom.  Similar experiments could also be carried out for $^{12}$Be and $^{20}$C.  $\rho_{cn}$ could also be determined by \emph{ab initio} approaches and then combined with halo-EFT~\cite{Zhang:2013kja,Hagen:2013jqa,Zhang:2014zsa}.

In this work the contribution of two-body $P$-wave interactions was not considered.  Such interactions can be added perturbatively as in Ref.~\cite{Margaryan:2015rzg} for the three-nucleon system.  However, for the two-neutron halos $^{6}$He and $^{17}$B resonant two-body $P$-wave interactions must be treated non-perturabitively~\cite{Ji:2014wta}.  This work also approximated all cores as spin zero, but  future work should consider arbitrary spin cores.  Further reduction of the theoretical error in halo-EFT will require a NNLO calculation.  However, at NNLO a new energy dependent three-body force, $h_{2}$, occurs that will require a new piece of three-body data.  The value for $h_{2}$ could be potentially fit to three-body data from \emph{ab inito} approaches or the asymptotic normalization of the halo nucleus wavefunction.  A future NNLO calculation will need to carefully consider appropriate renormalization conditions for $h_{2}$.

\acknowledgements{I would like to thank Daniel Phillips, Hans-Werner Hammer, Lucas Platter, and Bijaya Acharya for useful discussion during the course of this work.  In addition I would like to thank Daniel Phillips and Hans-Werner Hammer for valuable comments on the manuscript.  This material is based upon work supported by the U.S. Department of Energy, Office of Science, Office of Nuclear Physics, under Award Number DE-FG02-93ER40756.}
\appendix
\section{\label{app:Func}}

The matrix function $\boldsymbol{\mathcal{A}}_{m}^{(C)}(p,k,Q)$ for the diagram (a) contribution in Figs.~\ref{fig:HaloLOcharge} and \ref{fig:HaloNLOcharge} to the charge form factor is given by
\begin{align}
&\left[\boldsymbol{\mathcal{A}}_{m}^{(C)}(p,k,Q)\right]_{ij}=\frac{m_{n}^{2}A}{8\pi^{2}}
\int_{0}^{\Lambda}\!\!dq q^{2}\int_{0}^{1}\!\!dx\frac{1}{qQx}\left\{R_{i0}\left(E+\frac{qQx}{m_{n}A},k,d(q,Q,x)\right)\right.\\\nonumber
&\hspace{2cm}\times R_{0j}\left(E,d(q,Q,-x),p\right)\bar{D}_{nn}^{(m)}\left(E+\frac{qQx}{2m_{n}A}-\frac{Q^{2}}{4m_{n}A(2+A)},q\right)\\\nonumber
&\hspace{1cm}\left.-R_{i0}\left(E,k,d(q,Q,x)\right)R_{0j}\left(E-\frac{qQx}{m_{n}A},d(q,Q,-x),p\right)\right.\\\nonumber
&\left.\hspace{6cm}\times\bar{D}_{nn}^{(m)}\left(E-\frac{qQx}{2m_{n}A}-\frac{Q^{2}}{4m_{n}A(2+A)},q\right)\right\}\delta_{j1}\delta_{i1},
\end{align}
where $i,j=0,1$ label the matrix components, and $m=0$ ($m=1$) gives the LO contribution (NLO correction). $\boldsymbol{\mathcal{A}}_{m}^{(C)}(p,Q)$ the vector function is given by
\begin{align}
&\left[\boldsymbol{\mathcal{A}}_{m}^{(C)}(p,Q)\right]_{j}=\frac{m_{n}^{2}A}{8\pi^{2}}
\int_{0}^{\Lambda}\!\!dq q^{2}\int_{0}^{1}\!\!dx\frac{1}{qQx}\left\{R_{0j}\left(\frac{qQx}{m_{n}A}+E,d(q,Q,x),p\right)\right.\\\nonumber
&\hspace{2cm}\times\bar{D}_{nn}^{(m)}\left(E+\frac{qQx}{2m_{n}A}-\frac{Q^{2}}{4m_{n}A(2+A)},q\right)\\\nonumber
&\hspace{1cm}\left.-R_{0j}\left(E,d(q,Q,x),p\right)\bar{D}_{nn}^{(m)}\left(E-\frac{qQx}{2m_{n}A}-\frac{Q^{2}}{4m_{n}A(2+A)},q\right)\right\}\delta_{1j},
\end{align}
and the scalar function $\mathcal{A}_{m}^{(C)}(Q)$ by
\begin{align}
\mathcal{A}_{m}^{(C)}(Q)=\frac{m_{n}^{2}A}{8\pi^{2}}\int_{0}^{\Lambda}\!\!dq q^{2}\int_{0}^{1}\!\!dx\frac{1}{qQx}\left\{\bar{D}_{nn}^{(m)}\left(E+\frac{qQx}{2m_{n}A}-\frac{Q^{2}}{4m_{n}A(2+A)},q\right)\right.\\\nonumber
\left.-\bar{D}_{nn}^{(m)}\left(E-\frac{qQx}{2m_{n}A}-\frac{Q^{2}}{4m_{n}A(2+A)},q\right)\right\},
\end{align}
where
\begin{equation}
d(q,Q,x)=\sqrt{q^{2}+\frac{2qQx}{2+A}+\frac{Q^{2}}{(2+A)^{2}}}.
\end{equation}
For details of how to calculate the functions in this appendix consult Refs.~\cite{Hagen:2013xga,Vanasse:2015fph}.  The LO functions $\boldsymbol{\mathcal{A}}_{0}^{(C)}(p,k,Q)$, $\boldsymbol{\mathcal{A}}_{0}^{(C)}(p,Q)$, and $\mathcal{A}_{0}^{(C)}(Q)$ almost agree with the related functions of Hagen et \emph{al.}~\cite{Hagen:2013xga}, however, where we find the value $Q^{2}/(4m_{n}A(2+A))$ in the $nn$-dimer propagator they find $Q^{2}/(8m_{n}A)$.  Using their value for the $Q^{2}$ term we are able to reproduce the point charge radii given in their paper.  However, using their $Q^{2}$ value gives the wrong point charge radius in the equal mass and unitary limit, whereas our $Q^{2}$ value gives the correct point charge radius in this limit, given in Eq.~(\ref{eq:unitary}).

The matrix function $\boldsymbol{\mathcal{A}}_{m}^{(n)}(k,p,Q)$ contribution to diagram (a) in Figs.~\ref{fig:HaloLOmatter} and \ref{fig:HaloNLOmatter} for the neutron form factor is given by
\begin{align}
&\left[\boldsymbol{\mathcal{A}}_{m}^{(n)}(k,p,Q)\right]_{ij}=\frac{m_{n}^{2}}{8\pi^{2}}\int_{0}^{\Lambda}\!\!dq q^{2}\int_{0}^{1}\!\!dx\frac{1}{qQx}\left\{R_{1i}\left(\frac{qQx}{m_{n}}+E,d\left(q,\frac{1\!+\!A}{2}Q,x\right),k\right)\right.\\\nonumber
&\times R_{j1}\left(\vphantom{\frac{qQx}{m_{n}}}E,p,d\left(q,\frac{1\!+\!A}{2}Q,-x\right)\right)\bar{D}_{cn}^{(m)}\left(E+\frac{qQx}{2m_{n}}-\frac{(1\!+\!A)Q^{2}}{8m_{n}(2\!+\!A)},q\right)	\\\nonumber
&-R_{1i}\left(\vphantom{\frac{qQx}{m_{n}}}E,d\left(q,\frac{1\!+\!A}{2}Q,x\right),k\right)\\\nonumber
&\left.\times R_{j1}\left(-\frac{qQx}{m_{n}}+E,p,d\left(q,\frac{1\!+\!A}{2}Q,-x\right)\right)\bar{D}_{cn}^{(m)}\left(E-\frac{qQx}{2m_{n}}-\frac{(1\!+\!A)Q^{2}}{8m_{n}(2\!+\!A)},q\right)\right\},
\end{align}
the vector function $\boldsymbol{\mathcal{A}}_{m}^{(n)}(p,Q)$ by
\begin{equation}
\left[\boldsymbol{\mathcal{A}}_{m}^{(n)}(p,Q)\right]_{j}=0,
\end{equation}
and scalar function $\boldsymbol{\mathcal{A}}_{m}^{(n)}(Q)$ by
\begin{equation}
\mathcal{A}_{m}^{(n)}(Q)=0.
\end{equation}

Diagram (b) in Figs.~\ref{fig:HaloLOcharge} and~\ref{fig:HaloNLOcharge} for the charge form factor has the matrix function 
\begin{align}
&\left[\boldsymbol{\mathcal{B}}_{0}^{(C)}(p,k,Q)\right]_{ij}=\frac{m_{n}^{2}}{4}\int_{0}^{1}\!\!dx\int_{0}^{1}\!\!dy\int_{0}^{2\pi}\!\!d\phi\\\nonumber
&\times\frac{1}{k^{2}+p^{2}+\frac{2}{1+A}kp\theta(x,y,\phi)+\frac{Q^{2}}{(2+A)^{2}}+\frac{Q}{2+A}(2xk+\frac{2}{1+A}yp)-\frac{2m_{n}A}{(1+A)}E}\\\nonumber
&\times\frac{1}{k^{2}+p^{2}+\frac{2}{1+A}kp\theta(x,y,\phi)+\frac{Q^{2}}{(2+A)^{2}}-\frac{Q}{2+A}(2yp+\frac{2kx}{1+A})-\frac{2m_{n}A}{(1+A)}E}\delta_{j1}\delta_{i1},
\end{align}
while the neutron form factor diagram (b) and its time reversed version in Figs.~\ref{fig:HaloLOmatter} and \ref{fig:HaloNLOmatter} give the matrix function
\begin{align}
&\left[\boldsymbol{\mathcal{B}}_{0}^{(n)}(p,k,Q)\right]_{ij}=\frac{\sqrt{2}m_{n}^{2}(1+A)}{8A}\int_{0}^{1}\!\!dx\int_{0}^{1}\!\!dy\int_{0}^{2\pi}\!\!d\phi\\\nonumber
&\times\frac{1}{\frac{1+A}{2A}k^{2}+p^{2}+\frac{A(1+A)}{2(2+A)^{2}}Q^{2}+kp\theta(x,y,\phi)+\frac{Q}{2+A}((1+A)xk+Ayp)-m_{n}E}\\\nonumber
&\times\frac{1}{\frac{1+A}{2A}k^{2}+p^{2}+\frac{1}{(2+A)^{2}}Q^{2}+kp\theta(x,y,\phi)-\frac{Q}{2+A}(xk+2yp)-m_{n}E}\delta_{i1}\delta_{j0}\\\nonumber
&+\left(k\longleftrightarrow p\right) \delta_{i0}\delta_{j1}.
\end{align}
The $(k\longleftrightarrow p)$ represents the preceding term, but with $k$ and $p$ interchanged.  Due to time reversal symmetry the $\delta_{i0}\delta_{j1}$ term is equivalent to the $\delta_{i1}\delta_{j0}$ term.  Higher order corrections to the functions $\boldsymbol{\mathcal{B}}_{0}^{(X}(k,p,Q)$ do not exist.  Finally, the function $\theta(x,y,\phi)$ is the angle between vectors $\vect{k}$ and $\vect{p}$ and is defined as
\begin{equation}
\theta(x,y,\phi)=xy+\sqrt{1-x^{2}}\sqrt{1-y^{2}}\cos(\phi).
\end{equation}

The matrix function $\boldsymbol{\mathcal{C}}_{m}^{(C)}(p,k,Q)$ for diagram (c) of the charge form factor in Figs.~\ref{fig:HaloLOcharge} and~\ref{fig:HaloNLOcharge} is given by
\begin{align}
&\left[\boldsymbol{\mathcal{C}}_{m}^{(C)}(p,k,Q)\right]_{ij}=\frac{(1+A)m_{n}^{2}}{2\pi^{2}Q}\sum_{\alpha,\beta=0}^{1}\int_{0}^{\Lambda}\!\!dqq^{2}\int_{0}^{1}\!\!dx R_{i1}\left(E,k,d\left(q,\frac{Q}{2},x\right)\right)\\\nonumber
&\hspace{1cm}\times\bar{D}_{cn}^{(\alpha)}\left(E-\frac{qQx}{2m_{n}(1+A)}-\frac{Q^{2}}{8m_{n}(1+A)(2+A)},q\right)\\\nonumber
&\hspace{1cm}\times\arctan\left(\frac{\frac{Q}{2(2+A)}}{\sqrt{\frac{A(2+A)}{(1+A)^{2}}d^{2}\left(q,\frac{Q}{2},x\right)-\frac{2Am_{n}}{1+A}E}+\sqrt{\frac{A(2+A)}{(1+A)^{2}}d^{2}\left(q,\frac{Q}{2},-x\right)-\frac{2Am_{n}}{1+A}E}}\right)\\\nonumber
&\hspace{1cm}\times R_{1j}\left(E,d\left(q,\frac{Q}{2},-x\right),p\right)\bar{D}_{cn}^{(\beta)}\left(E+\frac{qQx}{2m_{n}(1\!+\!A)}-\frac{Q^{2}}{8m_{n}(1\!+\!A)(2\!+\!A)},q\right)\delta_{\alpha+\beta,m},
\end{align}
the vector contribution by
\begin{equation}
\left[\boldsymbol{\mathcal{C}}_{m}^{(C)}(k,Q)\right]_{j}=0,
\end{equation}
and the scalar contribution gives
\begin{equation}
\mathcal{C}_{m}^{(C)}(Q)=0.
\end{equation}
The type (c) diagram for the neutron form factor in Figs.~\ref{fig:HaloLOmatter} and~\ref{fig:HaloNLOmatter} has two contributions.  The first contribution is from a diagram with an intermediate $cn$-dimer and the second contribution has an intermediate $nn$-dimer.  Therefore, the neutron form factor matrix function $\boldsymbol{\mathcal{C}}_{m}^{(n)}(p,k,Q)$ will be split into
\begin{equation}
\boldsymbol{\mathcal{C}}_{m}^{(n)}(p,k,Q)=\boldsymbol{\mathcal{C}}_{m}^{(1;n)}(p,k,Q)+\boldsymbol{\mathcal{C}}_{m}^{(2;n)}(p,k,Q),
\end{equation}
the vector function split into
\begin{equation}
\boldsymbol{\mathcal{C}}_{m}^{(n)}(k,Q)=\boldsymbol{\mathcal{C}}_{m}^{(1;n)}(k,Q)+\boldsymbol{\mathcal{C}}_{m}^{(2;n)}(k,Q),
\end{equation}
and the scalar function split into
\begin{equation}
\mathcal{C}_{m}^{(n)}(Q)=\mathcal{C}_{m}^{(1;n)}(Q)+\mathcal{C}_{m}^{(2;n)}(Q),
\end{equation}
where the term with a $(1;n)$ ($(2;n)$) superscript refers to the diagram with an intermediate $cn$- ($nn$-) dimer.  The neutron form factor contribution from diagram (c) with an intermediate $cn$-dimer is similar to the (c) diagram for the charge form factor except the external current couples to the neutron instead of the core.  Thus the matrix function $\boldsymbol{\mathcal{C}}_{m}^{(1;n)}(p,k,Q)$ given by
\begin{align}
&\left[\boldsymbol{\mathcal{C}}_{m}^{(1;n)}(p,k,Q)\right]_{ij}=\frac{(1\!+\!A)^{2}m_{n}^{2}}{8\pi^{2}Q}\sum_{\alpha,\beta=0}^{1}\int_{0}^{\Lambda}\!\!dq q^{2}\int_{0}^{1}\!\! dx R_{1i}\left(E,d\left(q,\frac{Q}{2},-x\right),k\right)\\\nonumber
&\hspace{1cm}\times\bar{D}_{cn}^{(\alpha)}\left(E-\frac{Q^{2}}{8m_{n}(1\!+\!A)(2\!+\!A)}-\frac{qQx}{2m_{n}(1\!+\!A)},q\right)\\\nonumber
&\hspace{1cm}\times\bar{D}_{cn}^{(\beta)}\left(E-\frac{Q^{2}}{8m_{n}(1\!+\!A)(2\!+\!A)}+\frac{qQx}{2m_{n}(1\!+\!A)},q\right)\\\nonumber
&\hspace{1cm}\times\arctan\left(\frac{\frac{A}{1+A}Q}{\sqrt{\frac{A(2+A)}{(1+A)^{2}}d^{2}\left(q,\frac{Q}{2},-x\right)-\frac{2m_{n}A}{1+A}E}+\sqrt{\frac{A(2+A)}{(1+A)^{2}}d^{2}\left(q,\frac{Q}{2},x\right)-\frac{2m_{n}A}{1+A}E}}\right)\\\nonumber
&\hspace{1cm}\times R_{j1}\left(E,p,d\left(q,\frac{Q}{2},x\right)\right)\delta_{\alpha+\beta,1},
\end{align}
is the same as $\boldsymbol{\mathcal{C}}_{m}^{(C)}(p,k,Q)$ except for an overall constant and the $\arctan(...)$ term which is slightly different.  The vector function $\boldsymbol{\mathcal{C}}_{m}^{(1;n)}(k,Q)$ is given by
\begin{equation}
\left[\boldsymbol{\mathcal{C}}_{m}^{(1;n)}(k,Q)\right]_{j}=0,
\end{equation}
and the scalar function $\mathcal{C}_{m}^{(1;n)}(Q)$ by
\begin{equation}
\boldsymbol{\mathcal{C}}_{m}^{(1;n)}(Q)=0.
\end{equation}

For the neutron form factor the type (c) diagram with an intermediate $nn$-dimer of Figs.~\ref{fig:HaloLOmatter} and~\ref{fig:HaloNLOmatter} has the matrix function $\boldsymbol{\mathcal{C}}_{m}^{(2;n)}(p,k,Q)$ given by
\begin{align}
&\left[\boldsymbol{\mathcal{C}}_{m}^{(2;n)}(p,k,Q)\right]_{ij}=\frac{m_{n}^{2}}{2\pi^{2}Q}\sum_{\alpha,\beta=0}^{1}\int_{0}^{\Lambda}\!\!dqq^{2}\int_{0}^{1}\!\!dx \bar{D}_{nn}^{(\alpha)}\left(E+\frac{qQx}{4m_{n}}-\frac{AQ^{2}}{16m_{n}(2\!+\!A)},q\right)\\\nonumber
&\hspace{1cm}\times \bar{D}_{nn}^{(\beta)}\left(E-\frac{qQx}{4m_{n}}-\frac{AQ^{2}}{16m_{n}(2\!+\!A)},q\right)\\\nonumber
&\hspace{1cm}\times\arctan\left(\frac{\frac{1}{2}Q}{\sqrt{\frac{2+A}{4A}d^{2}\left(q,\frac{A}{2}Q,x\right)-m_{n}E}+\sqrt{\frac{2+A}{4A}d^{2}\left(q,\frac{A}{2}Q,-x\right)-m_{n}E}}\right)\\\nonumber
&\hspace{1cm}\times R_{0j}\left(E,d\left(q,\frac{A}{2}Q,-x\right),k\right)R_{i0}\left(E,p,d\left(q,\frac{A}{2}Q,x\right)\right)\delta_{j1}\delta_{i1}\delta_{\alpha+\beta,m},
\end{align}
the vector function
\begin{align}
&\left[\boldsymbol{\mathcal{C}}_{m}^{(2;n)}(k,Q)\right]_{j}=\frac{m_{n}^{2}}{\pi^{2}Q}\sum_{\alpha,\beta=0}^{1}\int_{0}^{\Lambda}\!\!dqq^{2}\int_{0}^{1}\!\!dx \bar{D}_{nn}^{(\alpha)}\left(E+\frac{qQx}{4m_{n}}-\frac{AQ^{2}}{16m_{n}(2\!+\!A)},q\right)\\\nonumber
&\hspace{1cm}\times \bar{D}_{nn}^{(\beta)}\left(E-\frac{qQx}{4m_{n}}-\frac{AQ^{2}}{16m_{n}(2\!+\!A)},q\right)\\\nonumber
&\hspace{1cm}\times\arctan\left(\frac{\frac{1}{2}Q}{\sqrt{\frac{2+A}{4A}d^{2}\left(q,\frac{A}{2}Q,x\right)-m_{n}E}+\sqrt{\frac{2+A}{4A}d^{2}\left(q,\frac{A}{2}Q,-x\right)-m_{n}E}}\right)\\\nonumber
&\hspace{1cm}\times R_{0j}\left(E,d\left(q,\frac{A}{2}Q,-x\right),k\right)\delta_{j1}\delta_{\alpha+\beta,m},
\end{align}
and the scalar function
\begin{align}
&\mathcal{C}_{m}^{(2;n)}(Q)=\frac{m_{n}^{2}}{2\pi^{2}Q}\sum_{\alpha,\beta=0}^{1}\int_{0}^{\Lambda}\!\!dqq^{2}\int_{0}^{1}\!\!dx \bar{D}_{nn}^{(\alpha)}\left(E+\frac{qQx}{4m_{n}}-\frac{AQ^{2}}{16m_{n}(2\!+\!A)},q\right)\\\nonumber
&\times \bar{D}_{nn}^{(\beta)}\left(E-\frac{qQx}{4m_{n}}-\frac{AQ^{2}}{16m_{n}(2\!+\!A)},q\right)\\\nonumber
&\times\arctan\left(\frac{\frac{1}{2}Q}{\sqrt{\frac{2+A}{4A}d^{2}\left(q,\frac{A}{2}Q,x\right)-m_{n}E}+\sqrt{\frac{2+A}{4A}d^{2}\left(q,\frac{A}{2}Q,-x\right)-m_{n}E}}\right)\delta_{\alpha+\beta,m}.
\end{align}
The $\arctan(...)$ function for all (c) type diagrams comes from analytically solving the two-body bubble sub-diagram.  Our matrix functions $\boldsymbol{\mathcal{B}}_{0}^{C}(p,k,Q)$ and $\boldsymbol{\mathcal{C}}_{0}^{(C)}(p,k,Q)$ agree with the associated functions of Hagen et \emph{al.}~\cite{Hagen:2013xga}.

At NLO the diagram (d) contribution to the charge form factor in Fig.~\ref{fig:HaloNLOcharge} gives the matrix function
\begin{align}
&\left[\boldsymbol{\mathfrak{D}}_{1}^{(C)}(p,k,Q)\right]_{ij}=-\frac{m_{n}^{2}}{16\pi^{2}}\frac{Z_{cn}-1}{2\gamma_{cn}}\int_{0}^{\Lambda}\!\!dqq^{2}\int_{0}^{1}\!\!dx R_{i1}\left(E,k,d\left(q,\frac{Q}{2},x\right)\right)\\\nonumber
&\hspace{2cm}\times\bar{D}_{cn}^{(0)}\left(E-\frac{qQx}{2m_{n}(1\!+\!A)}-\frac{Q^{2}}{8m_{n}(1\!+\!A)(2\!+\!A)},q\right)\\\nonumber
&\hspace{2cm}\times R_{1j}\left(E,d\left(q,\frac{Q}{2},-x\right),p\right)\bar{D}_{cn}^{(0)}\left(E+\frac{qQx}{2m_{n}(1\!+\!A)}-\frac{Q^{2}}{8m_{n}(1\!+\!A)(2\!+\!A)},q\right),
\end{align}
vector function
\begin{equation}
\left[\boldsymbol{\mathfrak{D}}_{1}^{(C)}(k,Q)\right]_{j}=0,
\end{equation}
and scalar function
\begin{equation}
\mathfrak{D}_{1}^{(C)}(Q)=0.
\end{equation}
These functions are entirely analogous to the functions for the charge form factor contribution to diagram (c).  This is because diagram (d) is essentially diagram (c) with the two-body sub-diagram replaced with a direct coupling to the gauged $cn$-dimer.

The diagram (d) contribution to the neutron form factor is split up into two parts in complete analogy to the diagram (c) contribution.  Diagram (d) with an intermediate $cn$-dimer gives the matrix function
\begin{align}
&\left[\boldsymbol{\mathfrak{D}}_{1}^{(1;n)}(p,k,Q)\right]_{ij}=-\frac{m_{n}^{2}}{16\pi^{2}}\frac{Z_{cn}-1}{2\gamma_{cn}}\int_{0}^{\Lambda}\!\!dq q^{2}\int_{0}^{1}\!\! dx R_{1i}\left(E,d\left(q,\frac{Q}{2},-x\right),k\right)\\\nonumber
&\hspace{1cm}\times\bar{D}_{cn}^{(0)}\left(E-\frac{Q^{2}}{8m_{n}(1\!+\!A)(2\!+\!A)}-\frac{qQx}{2m_{n}(1\!+\!A)},q\right)\\\nonumber
&\hspace{1cm}\times\bar{D}_{cn}^{(0)}\left(E-\frac{Q^{2}}{8m_{n}(1\!+\!A)(2\!+\!A)}+\frac{qQx}{2m_{n}(1\!+\!A)},q\right)\\\nonumber
&\hspace{1cm}\times R_{j1}\left(E,p,d\left(q,\frac{Q}{2},x\right)\right),
\end{align}
vector function
\begin{equation}
\left[\boldsymbol{\mathfrak{D}}_{1}^{(1;n)}(k,Q)\right]_{j}=0,
\end{equation}
and scalar function
\begin{equation}
\boldsymbol{\mathfrak{D}}_{1}^{(1;n)}(Q)=0.
\end{equation}
Finally, diagram (d) with an intermediate $nn$-dimer has the matrix function
\begin{align}
&\left[\boldsymbol{\mathfrak{D}}_{1}^{(2;n)}(k,p,Q)\right]_{ij}=-\frac{m_{n}^{2}}{8\pi^{2}}\frac{Z_{nn}-1}{2\gamma_{nn}}\int_{0}^{\Lambda}\!\!dqq^{2}\int_{0}^{1}\!\!dx \bar{D}_{nn}^{(0)}\left(E+\frac{qQx}{4m_{n}}-\frac{AQ^{2}}{16m_{n}(2\!+\!A)},q\right)\\\nonumber
&\hspace{1cm}\times \bar{D}_{nn}^{(0)}\left(E-\frac{qQx}{4m_{n}}-\frac{AQ^{2}}{16m_{n}(2\!+\!A)},q\right)\\\nonumber
&\hspace{1cm}\times R_{0j}\left(E,d\left(q,\frac{A}{2}Q,-x\right),k\right)R_{i0}\left(E,p,d\left(q,\frac{A}{2}Q,x\right)\right)\delta_{j1}\delta_{i1},
\end{align}
vector function
\begin{align}
&\left[\boldsymbol{\mathfrak{D}}_{1}^{(2;n)}(k,Q)\right]_{j}=-\frac{m_{n}^{2}}{4\pi^{2}}\frac{Z_{nn}-1}{2\gamma_{nn}}\int_{0}^{\Lambda}\!\!dqq^{2}\int_{0}^{1}\!\!dx \bar{D}_{nn}^{(0)}\left(E+\frac{qQx}{4m_{n}}-\frac{AQ^{2}}{16m_{n}(2\!+\!A)},q\right)\\\nonumber
&\hspace{1cm}\times \bar{D}_{nn}^{(0)}\left(E-\frac{qQx}{4m_{n}}-\frac{AQ^{2}}{16m_{n}(2\!+\!A)},q\right)\\\nonumber
&\hspace{1cm}\times R_{0j}\left(E,d\left(q,\frac{Q}{2},-x\right),k\right)\delta_{j1},
\end{align}
and scalar function
\begin{align}
&\mathfrak{D}_{1}^{(2;n)}(Q)=-\frac{m_{n}^{2}}{8\pi^{2}}\frac{Z_{nn}-1}{2\gamma_{nn}}\int_{0}^{\Lambda}\!\!dqq^{2}\int_{0}^{1}\!\!dx \bar{D}_{nn}^{(0)}\left(E+\frac{qQx}{4m_{n}}-\frac{AQ^{2}}{16m_{n}(2\!+\!A)},q\right)\\\nonumber
&\times \bar{D}_{nn}^{(0)}\left(E-\frac{qQx}{4m_{n}}-\frac{AQ^{2}}{16m_{n}(2\!+\!A)},q\right).
\end{align}
These functions are again completely analogous to their type (c) diagram counterparts.  Diagram (e) in Fig.~\ref{fig:HaloNLOcharge} for the charge form factor is subtracted from diagram (a) and its time reversed version in Fig.~\ref{fig:HaloNLOcharge} to avoid double counting.  Therefore, the contribution from diagram (e) in Fig.~\ref{fig:HaloNLOcharge} has been included in the functions $\boldsymbol{\mathcal{A}}_{1}^{(C)}(...)$.  The same procedure is carried out for diagram (e) for the neutron form factor in Fig.~\ref{fig:HaloNLOmatter}.

%\bibliography{HaloSwave}
%\bibliographystyle{apsrev} 

\end{document}